\documentclass[aps,pra,twocolumn,showpacs,superscriptaddress,nofootinbib]{revtex4-1}

\usepackage{graphicx}
\usepackage[utf8]{inputenc}
\usepackage[normalem]{ulem}
\usepackage{fontenc}
\usepackage{fontenc}
\usepackage{mathtools}
\usepackage{amsmath}
\usepackage{amssymb}
\usepackage{amsthm}
\usepackage{amsmath,amssymb,amsthm,graphicx, array,epsfig, exscale, dsfont, amsfonts, verbatim, fancyhdr, natbib, bbm,mathrsfs,textcomp,listings, epsfig,bm}
\usepackage{slashbox}
\usepackage{color}
\usepackage{txfonts} 

\newcommand{\centre}[2]{\multispan{#1}{\hfill #2\hfill}}
\newcommand{\crule}[1]{\multispan{#1}{\hspace*{\tabcolsep}\hrulefill
  \hspace*{\tabcolsep}}}
\newcommand{\one}{\mbox{$1 \hspace{-1.0mm}  {\bf l}$}}
\newcommand{\bra}[1]{\left\langle{#1}\right\vert}
\newcommand{\ket}[1]{\left\vert{#1}\right\rangle}

\newcommand{\pdark}{p_{\mathrm{dark}}}
\newcommand{\pD}{\pdark}
\newcommand{\etaD}{\eta_\mathrm{d}}
\newcommand{\nl}{N}

\newcommand{\etaM}{\eta_{m}}
\newcommand{\pdis}{P_D}

\date{\today}

\newcommand{\eci}[1]{e_{#1}}

\newcommand{\rrep}{R_{\mathrm{REP}}}
\newcommand{\rraw}{R_{\mathrm{raw}}}
\newcommand{\rsift}{R_{\mathrm{sift}}}
\newcommand{\etaT}[1]{\eta_{t}\left(#1\right)}
\newcommand{\proj}[1]{\left\vert{#1}\right\rangle \left\langle{#1}\right\vert}
\newcommand{\tr}{\mathrm{tr}}

\newcommand{\sref}[1]{Sec.~\ref{#1}}
\newcommand{\fref}[1]{Fig.~\ref{#1}}
\newcommand{\Fref}[1]{Figure~\ref{#1}}
\newcommand{\eref}[1]{Eq.~\eqref{#1}}
\newcommand{\Tref}[1]{Table~\ref{#1}}
\newcommand{\tref}[1]{Tab.~\ref{#1}}
\newcommand{\etal}{\emph{et al.~}}

\newcommand{\duaff}{Institute for Theoretical Physics III, Heinrich-Heine-Universit\"at D\"usseldorf, Universitätsstr. 1, 40225 D\"usseldorf, Germany}

\newcommand{\erlaff}{Optical Quantum Information Theory Group, Max Planck Institute for the Science of Light, Günther-Scharowsky-Str. 1/Bau 24, 91058 Erlangen, Germany}
\newcommand{\erluniaff}{Institute of Theoretical Physics I, Universität Erlangen-Nürnberg, Staudtstr. 7/B2, 91058 Erlangen, Germany}
\newcommand{\mainzaff}{Institute of Physics, Johannes-Gutenberg Universit\"at Mainz, Staudingerweg 7, 55128 Mainz, Germany}

\begin{document}

\title{Quantum repeaters and quantum key distribution: analysis of secret key rates}

\author{Silvestre Abruzzo}
\email{abruzzo@thphy.uni-duesseldorf.de}
\affiliation{\duaff}

\author{Sylvia Bratzik}
\affiliation{\duaff}

\author{Nadja K Bernardes}
\affiliation{\erlaff}
\affiliation{\erluniaff}

\author{Hermann Kampermann}
\affiliation{\duaff}

\author{Peter van Loock}
\affiliation{\erlaff}
\affiliation{\erluniaff}
\affiliation{\mainzaff}

\author{Dagmar Bru{\ss}}
\affiliation{\duaff}

\begin{abstract}
We analyze various prominent quantum repeater protocols in the context of long-distance quantum key distribution. These protocols are the original quantum repeater proposal by Briegel, Dür, Cirac and Zoller, the so-called hybrid quantum repeater using optical coherent states dispersively interacting with atomic spin qubits, and the Duan-Lukin-Cirac-Zoller-type repeater using atomic ensembles together with linear optics and, in its most recent extension, heralded qubit amplifiers. For our analysis, we investigate the most important experimental parameters of every repeater component and find their minimally required values for obtaining a nonzero secret key. Additionally, we examine in detail the impact of device imperfections on the final secret key rate and on the optimal number of rounds of distillation when the entangled states are purified right after their initial distribution.
\end{abstract}


\maketitle

\section{Introduction}
Quantum communication is one of the most exciting and well developed areas of quantum information. Quantum key distribution (QKD) is a sub-field, where two parties, usually called Alice and Bob, want to establish a secret key. For this purpose,  typically, they perform some quantum operations on two-level systems, the qubits, which, for instance, can be realized by using polarized photons.  \cite{ekert1991quantum,PhysRevA.84.022325,PhysRevA.84.010304,PhysRevLett.105.070501,PhysRevLett.98.230501}.

Photons naturally have a long decoherence time and hence could be transmitted over long distances. Nevertheless, recent experiments show that QKD so far is limited to about 150 km
\cite{Scarani:2009}, due to losses in the  optical-fiber channel. Hence, the concept of quantum relays and repeaters was developed \cite{briegel_quantum_1998, RevModPhys.74.145, gisinrelay, PhysRevLett.92.047904,PhysRevA.65.052310}. These aim at entangling qubits over long distances by means of entanglement swapping and entanglement distillation. There exist various proposals for an experimental implementation, such as
 those based upon atomic ensembles and single-rail entanglement \cite{duan_long-distance_2001}, the hybrid quantum repeater \cite{van_loock_hybrid_2006}, the ion-trap quantum repeater \cite{PhysRevA.79.042340}, repeaters based on  deterministic Rydberg gates \cite{PhysRevA.81.052329, PhysRevA.81.052311}, and repeaters based on nitrogen-vacancy (NV) centers in diamond \cite{childress_fault-tolerant_2005}.

In this paper, we analyze the performance of quantum repeaters within a QKD set-up, for calculating secret key rates as a function of the relevant experimental parameters. Previous investigations on long-distance QKD either consider quantum relays \cite{gisinrelay,PhysRevA.65.052310,scherer_long-distance_2011}, which only employ entanglement swapping
 without using quantum memories or entanglement distillation, or,  like the works in \cite{razavi_quantum_2010,amirloo_quantum_2010}, they  exclusively refer to the original Duan-Lukin-Cirac-Zoller (DLCZ) quantum repeater \cite{duan_long-distance_2001}. Finally, in \cite{2012arXiv1210.8042L} the authors analyze a variation of the DLCZ protocol \cite{2007PhRvA.76e0301S} where they consider at most one repeater station. Here, our aim is to quantify the influence of characteristic experimental parameters on the secret key rate for  three different repeater schemes, namely the original quantum repeater protocol \cite{briegel_quantum_1998}, the hybrid quantum repeater \cite{van_loock_hybrid_2006}, and a  recent variation of the DLCZ-repeater \cite{minar}. We investigate the minimally required parameters that allow a non-zero secret key rate.  In order to reduce the complexity of  the
 full repeater
protocol, we consider  entanglement distillation  only directly after the  initial entanglement distribution. Within this scenario, we investigate also the optimal number of
distillation rounds for a wide range of parameters. The influence of distillation  during later stages of the repeater, as well as the comparison between different distillation protocols, will be studied elsewhere \cite{Bratzik2012}.

This manuscript is organized as follows: In \sref{sec:genfrac} we present a description of the relevant parameters of a quantum repeater, as well as the main tools for analyzing  its performance for QKD. This section should also provide a  general framework  for analyzing other existing
quantum repeater protocols, and for studying the performance and the potential of new protocols.  Sections \ref{sec:briegel}, \ref{sec:Hybrid}, and \ref{sec:atomicensenbles} investigate  long-distance QKD protocols for three different quantum repeater schemes; these  sections can be read independently. Section~\ref{sec:briegel}  is devoted to the original proposal  for a quantum repeater \cite{briegel_quantum_1998}, section~\ref{sec:Hybrid} analyzes the hybrid quantum repeater \cite{van_loock_hybrid_2006}, and finally, in section~\ref{sec:atomicensenbles}, we investigate quantum repeaters with atomic ensembles \cite{duan_long-distance_2001}. The conclusion will be given in section~\ref{sec:conclusions},
 and more details on the calculations will be presented in the appendix.

\section{General framework}
\label{sec:genfrac}

\subsection{Quantum repeater}
The purpose of this section is to provide a general framework that describes formally the theoretical analysis of a quantum repeater.

\subsubsection{The protocol} 
\begin{figure}[h]
\centering
\includegraphics[width=8cm, clip]{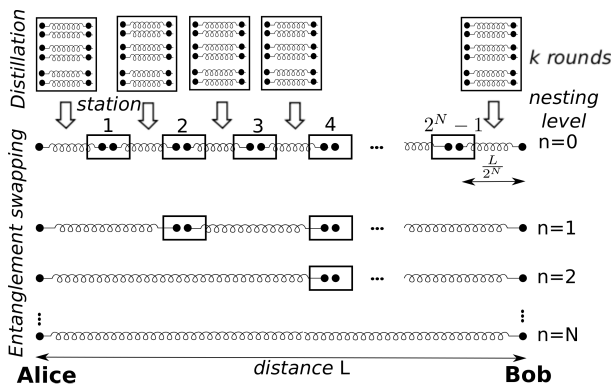}
\caption{Scheme of a generic quantum repeater protocol. We adopt the nested protocol proposed in \cite{briegel_quantum_1998}. The distance between Alice and Bob is $L$, which is divided in $2^{N}$ segments, each having the length $L_0:=L/2^{N}$. The parameter $n$ describes the different nesting levels, and the value $N$  represents the maximum nesting level. In this paper, we consider quantum repeaters where distillation is performed  exclusively before the first entanglement  swapping step. The number of distillation rounds is denoted by $k$.
\label{fig:qr}}
\end{figure}

Let $L$ be the distance between the two parties Alice and Bob who wish to share an entangled state. A quantum repeater \cite{briegel_quantum_1998} consists of a chain of $2^N$ segments of fundamental length $L_0:=L/2^N$ and $2^N-1$ repeater stations which are placed at the intersection points between two segments (see \fref{fig:qr}).  Each repeater station  is equipped with quantum memories and local quantum processors to perform entanglement swapping and, in general, also entanglement distillation. In consecutive \textit{nesting levels}, the  distances over which the entangled  states are shared will be doubled. The parameter $N$ is the \textit{maximal nesting level}.

The protocol starts by creating entangled states in all segments, i.e., between two quantum memories  over distance $L_0$. After that, if necessary,  entanglement distillation is performed.  This distillation is a probabilistic process which requires sufficiently many initial pairs  shared over distance $L_0$.  As a next step, entanglement swapping is performed  at the corresponding repeater stations in order to  connect two adjacent entangled pairs and thus  gradually extend the entanglement. In those protocols where entanglement swapping is a probabilistic process, the whole quantum repeater protocol is performed in a recursive way as shown in \fref{fig:qr}.  Whenever the swapping is deterministic (i.e., it never fails), then all swappings can be executed simultaneously, provided that no further probabilistic entanglement distillation steps are to be incorporated at some intermediate nesting levels for enhancing the fidelities. Recall that in
the present work, we do not include such intermediate distillations in order to keep the experimental requirements as low as possible. At the same time it allows us to find analytical rate formulas with no need for numerically optimizing the distillation-versus-swapping scheduling in a fully nested quantum repeater.


\subsubsection{\label{subsec:parameters}Building blocks of the quantum repeater and their imperfections}

In this section we describe a model of the imperfections for the main building blocks of a quantum repeater. In an experimental set-up more imperfections than those considered in this model may affect the devices. However,  most of them can be incorporated into our model. We point out that if not all possible imperfections are included, the resulting curves for the figure of merit (throughout this paper: the secret key rate) can be interpreted as an upper bound for a given repeater protocol.

\paragraph{Quantum channel}
 Let us consider photons (in form of single- or multi-photon pulses) traveling through optical fibers.

 Photon losses are the main source of imperfection. Other imperfections like birefringence are negligible in our context \cite{RevModPhys.74.145, sangouard_quantum_2011}. Losses scale exponentially with the length $\ell$, i.e., the transmittivity is given by \cite{RevModPhys.74.145}
\begin{equation}
\etaT{\ell}:=10^{-\frac{\alpha_{att} \ell}{10}},
\label{eq:etaT}
\end{equation}
where $\alpha_{att}$ is the attenuation coefficient given in dB/km. The lowest attenuation is achieved in the telecom wavelength range around 1550 nm and  it corresponds to  $\alpha_{att}=0.17$ dB/km. This attenuation will also be used throughout the paper. Note that other types of quantum channels, such as free space, can be treated in an equivalent way (see e.g. \cite{tatarski1961wave}).
Further note that besides losses, the effect of the quantum channel can be incorporated into the form of the initial state shared between the connecting repeater stations.

\paragraph{Source of entanglement}
The purpose of a source is to create entanglement between quantum memories  over distance $L_0$.  An ideal source produces maximally entangled Bell states (see below) on demand. In practice,  however, the created state may not be maximally entangled and may be produced in a probabilistic way. We denote by $\rho_0$ a state shared between two quantum memories over the elementary distance $L_0$ and by $P_0$ the total probability to generate and distribute this state. This probability would contain any finite local state-preparation probabilities before the distribution, the effect of channel losses, and
the success probabilities of other processes, such as the conditioning on a desired initial state $\rho_0$ after the state distribution over $L_0$.

For improving the scaling over the total distance $L$ from exponential to sub-exponential, it is necessary to have a heralded creation and storage of $\rho_0$.  How this heralding is implemented depends on the  particular protocol and it usually involves a form of post-processing, e.g. conditioning the state on a specific pattern of detector clicks. This can also be a finite postselection window of quadrature values in homodyne detection. However,
in the present work, the measurements employed in all protocols considered here are either photon-number measurements or Pauli measurements on memory qubits.

\paragraph{Detectors}
We will consider photon-number resolving detectors (PNRD) which can be described by  a positive-operator valued measure (POVM) with elements \cite{kok_introduction_2010}
\begin{equation}
\label{eq:POVMPNRD}
 \Pi^{(n)}:=\etaD^{n}\sum_{m=0}^{\infty}{n+m \choose n} (1-\etaD)^m\ket{n+m}\bra{n+m}.
\end{equation}
 Here, $\Pi^{(n)}$ is the element of the POVM related to the detection of $n$ photons, $\etaD$ is the efficiency of the detector, and $\ket{n+m}$ is a state of $(n+m)$-photons. In the POVM above, we have neglected dark counts; we have shown analytically for those protocols considered in this paper that realistic dark counts of the order of $10^{-5}$ are negligible [see Appendix~\ref{app:general-type}, below Eq.~\eqref{eq:briegpes}, for the proof].
 Note that our analysis could also be extended to threshold detectors, by replacing the corresponding POVM (see e.g.\ \cite{kok_introduction_2010}) in our formulas.

\paragraph{Gates}
Imperfections of gates also depend on the particular quantum repeater implementation. Such imperfections are e.g.\ described in \cite{gilchrist2005distance}. In our analysis, we will characterize them using  the gate quality which will be denoted by $p_G$ (see \eref{eq:p2} and \eref{gateerror}). 

\paragraph{Quantum memories}
Quantum memories are a crucial part of a quantum repeater. A complete characterization of imperfections of quantum memories is beyond the purpose of this paper (see \cite{qmemrev} for a recent review). Here we account for memory errors by using a fixed time-independent quantum memory efficiency $\eta_{m}$ when appropriate. This is the probability that a photon is released when a reading signal is applied to the quantum memory, or, more generally, the probability that an initial qubit state is still intact after write-in, storage, and read-out. We discuss the role of $\eta_{m}$ only for the quantum repeater with atomic ensembles (see section~\ref{sec:atomicensenbles}).
\paragraph{Entanglement distillation}
As mentioned  before, throughout this work we only consider distillation at the beginning of  each repeater protocol. Entanglement distillation is a probabilistic process  requiring local multi-qubit gates  and classical communication. In this paper, we consider the protocol by Deutsch \etal \cite{deutsch1996quantum}. This protocol performs  especially well when there are different types of errors (e.g. bit flips and phase flips). However, depending on the  particular form of the initial state and on the particular quantum repeater protocol, other distillation  schemes may perform better (see \cite{Bratzik2012} for a detailed discussion). The Deutsch \etal protocol starts with $2^k$ pairs and after $k$ rounds, it produces one entangled pair with higher fidelity  than at the beginning.  Every round requires two Controlled Not (CNOT),
each performed on two  qubits at the same  repeater station, and  projective measurements with post-selection.

Distillation has two main sources of  errors: imperfect quantum gates which  no longer permit to  achieve the ideal fidelity, as well as imperfections of the quantum memories and the detectors,  decreasing the success probability. We denote the success probability in the $i$-th distillation round by $P_{D}[i]$.

We study entanglement distillation for the original quantum repeater protocol (section~\ref{sec:briegel}) and the hybrid quantum repeater (section~\ref{sec:Hybrid}). For the quantum repeater with atomic ensembles (section~\ref{sec:atomicensenbles}), we do not consider any additional distillation on two or more initial memory pairs.

\paragraph{Entanglement swapping}
 In order to extend the initial distances of the shared entanglement,
entanglement swapping  can be achieved through a Bell measurement  performed at the  corresponding  stations between two adjacent segments.  Such a Bell measurement can be, in principle,  realized using a CNOT gate and suitable projection measurements on the corresponding quantum memories \cite{Zukowski1993}. An alternative implementation of the Bell measurement uses photons released from the quantum memories and linear optics \cite{Pan1998}.  The latter technique is probabilistic, but typically much less demanding from  an experimental point of view.

We  should emphasize that the  single-qubit rotation  depending on the result of the Bell measurement, as generally needed to complete the entanglement swapping step, is not necessary when the final state is used for  QKD applications. In fact, it simply corresponds to  suitable bit flip operations on the outcomes of the QKD measurements, i.e.,  the effect of that single-qubit rotation can be included into the classical post-processing.

Imperfections of entanglement swapping are characterized by the imperfections of the gates (which introduce noise and therefore a decrease in fidelity) and by the imperfections of the measurement process, caused by imperfect quantum memories and imperfect detectors. We denote the probability that entanglement swapping is successful in the $n$-th nesting level by $P_{\rm ES}^{(n)}$.

\paragraph{Other imperfections}
Other imperfections which are not explicitly  considered in this paper but which are likely to be present in a real experiment include imperfections of the interconversion process,  fluctuations of the quantum channel, fiber coupling losses and passive losses of optical elements (see \cite{sangouard_quantum_2011} and reference therein for additional details). These imperfections can be accounted for by a suitable adjustment of the relevant parameters in our model.

\subsubsection{\label{subsec:gen}Generation rate of long-distance entangled pairs}
In order to evaluate the performance of a quantum repeater protocol it is necessary to assess how many entangled pairs across distance $L$ can be generated per second.

\begin{figure}
 \centering
 \includegraphics[width=8cm, clip]{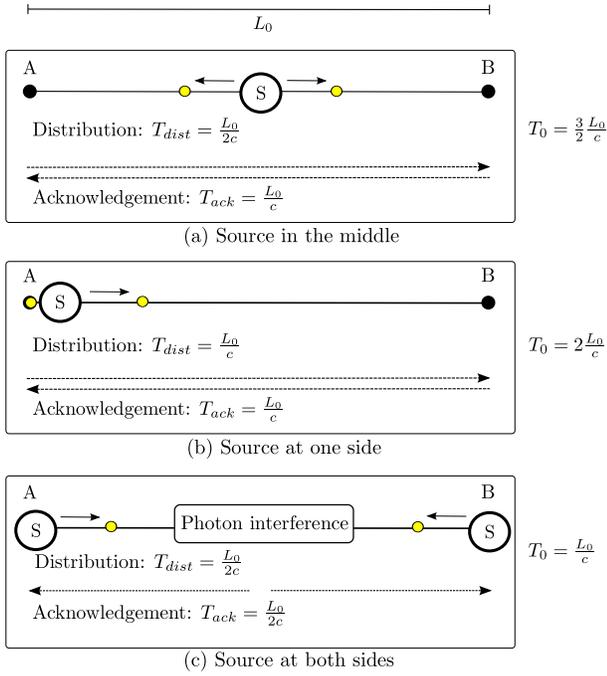}
 \caption{The fundamental time for different models of entanglement generation  and distribution. The source (S) that produces the initial entangled states is either placed in the middle (a), at one side (b), or at both sides (c). In the latter case, photons are emitted from a source and interfere in the middle (see \cite{Cabrillo1999,Feng2003}).}
 \label{fig:comm_time}
\end{figure}

A relevant unit of time is the \textit{fundamental time} needed to communicate the successful distribution of an elementary entangled pair  over distance $L_0$, which is given by:
\begin{equation}
\label{eq:fundtime}
T_{0}:=\frac{\beta L_0}{c},
\end{equation}
where $c=2\cdot10^5$ km/s is the speed of light in the  fiber channel (see e.g.\ \cite{sangouard_quantum_2011}) and $\beta$ is a factor depending on the type of entanglement distribution. Note that here we have neglected the additional local times needed for preparing and manipulating the physical systems at each repeater station. \Fref{fig:comm_time} shows three different possibilities how to model  the initial entanglement distribution. The fundamental time $T_0$ consists of the time to distribute the  photonic signals, $T_{dist}$,  and the time of acknowledgment, $T_{ack}$, which  all together can be different for the three cases shown.

Throughout the paper, we denote the average number of final entangled pairs produced in the repeater per second by $\rrep$. We emphasize that  regarding any figures and plots, for each protocol, we are interested in the  consumption of time rather than spatial memories. Thus, if one wants to compare different set-ups for the same number of  spatial memories, one has to rescale the rates such that the number of memories becomes equal. For example, in order to compare a  protocol without distillation with another one with $k$ rounds of distillation, one has to divide the  rates for the case with distillation by $2^k$ (as we need two initial pairs  to obtain one distilled pair in every round).

In the literature, two different upper bounds on the entanglement generation rate $\rrep$ are known. In the case of deterministic entanglement swapping ($P_{ES}^{(n)}=1$) we have \cite{NadjaHybrid}

\begin{equation}
\rrep^{\rm det}=\left(T_0 {Z_{N}(P_{L_0}[k])}\right)^{-1},
\label{eq:zn:approx}
\end{equation}
with $P_{L_0}[i]$ being a recursive probability depending on the rounds of distillation $i$ as follows \cite{NadjaHybrid}
\begin{eqnarray}
P_{L_0}[i=0]&=&P_0,\\
P_{L_0} [i>0]&=&\frac{\pdis[i]}{Z_{1}(P_{L_0}[i-1])}.
\label{eq:PL0}
\end{eqnarray}
We remind the reader that $P_D[i]$ is the success probability in the $i$-th distillation round.
Here,
\begin{equation}
\label{eq:zn}
 Z_{N}(P_0):=\sum_{j=1}^{2^N}{2^N \choose j}\frac{(-1)^{j+1}}{1-(1-P_0)^j}
\end{equation}
is the average number of attempts to connect $2^N$ pairs, each generated with probability $P_0$.

In the case of probabilistic entanglement swapping,  probabilistic entanglement distillation, and $P_0<<1$, we find an upper bound on the entanglement generation rate:
\begin{equation}
\label{eq:avgngen}
 \rrep^{\rm prob}= \frac{1}{T_0}\left(\frac{2}{3 a}\right)^{N+k} P_{0}P_{ES}^{(1)}P_{ES}^{(2)}...P_{ES}^{(\nl)}\prod_{i=1}^{k}\pdis[i],
\end{equation}
with $a\leq\frac{2}{3}P_{L_0}[k] Z_1(P_{L_0}[k])$.
Our derivation is given in App.~\ref{sec:app:general}. For the plots we bound  $a$ according to the occuring parameters, typically $a$ is close to one which corresponds to the approximate formula given in \cite{sangouard_quantum_2011} for the case when there is no distillation.

Equations~\eqref{eq:zn:approx} and \eqref{eq:avgngen} should be interpreted as a limiting upper bound on the repeater  rate,  due to the minimal time needed for communicating the quantum and classical signals. For this minimal time , we consider explicitly only those communication times for initially generating entanglement, but not those for entanglement swapping and entanglement distillation.

\subsection{Quantum key distribution (QKD)}
\label{sec:QKD}
\subsubsection*{The QKD protocol}
\label{subsec:qkd}

\begin{figure}[h]
\centering
\includegraphics[width=8cm, clip]{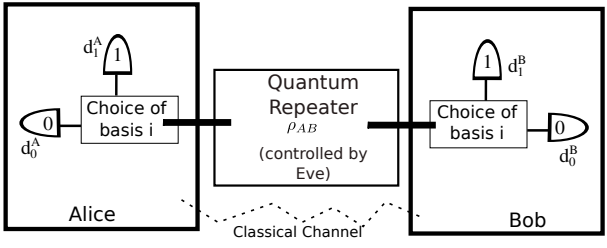}
\caption{Scheme of quantum key distribution. The state $\rho_{AB}$  is produced using a quantum repeater. Alice and Bob  locally rotate  this state in a measurement basis and then they perform the measurement. The detectors are denoted by $d_0^A,d_1^A,d_0^B,d_1^B$ and to each detector click a classical outcome is assigned.}
\label{fig:QKDSetup}
\end{figure}

In \fref{fig:QKDSetup} a general quantum key distribution set-up is shown.
For long-distance QKD, Alice and Bob will generate entangled pairs using the quantum repeater
protocol. For the security analysis of the whole repeater-based QKD scheme, we assume that  a potential eavesdropper (Eve) has
complete control of the repeater stations, the quantum channels connecting them, and the classical channels used for communicating the measurement outcomes for entanglement swapping and distillation (see figure~\ref{fig:QKDSetup}).
The QKD protocol itself starts with Alice and Bob performing
measurements on their shared, long-distance entangled pairs (see figure~\ref{fig:QKDSetup}). For this purpose, they would both independently choose a certain measurement from a given set of measurement settings.
 The next step is the classical post-processing and for this an authenticated channel is necessary.
First, Alice and Bob discard those measurement outcomes where their choice of the setting did not coincide (sifting), thus obtaining a raw key associated with a \textit{raw key rate}. They proceed by comparing publicly a small subset of outcomes (parameter estimation). From this subset, they can estimate the \emph{quantum bit error rate} (QBER), which corresponds to the fraction of uncorrelated bits. If the QBER is below a certain threshold, they apply an error correction protocol and privacy amplification in order to shrink the eavesdropper's information about the secret key (for more details, see e.g. \cite{Renner2008}).

Various QKD protocols exist in the literature. Besides the original QKD protocol by Bennett and Brassard  from 1984, the so-called BB84-protocol \cite{bennett1984quantum}, the first QKD protocol based upon entanglement was the Ekert protocol \cite{ekert1991quantum}. Shortly thereafter the relation of the Ekert protocol to the BB84-protocol was found \cite{bennett_quantum_1992}. Another protocol which can also be applied in  entanglement-based QKD is the six-state protocol \cite{bruss1998optimal, 6stategisin}.

\subsubsection{The quantum bit error rate (QBER)}

In order to evaluate the performance of a QKD protocol, it is necessary to determine the quantum bit error rate. This is the fraction of discordant outcomes when Alice and Bob compare a small amount of outcomes taken from a specified measurement basis. This measurement can be modelled by means of four detectors (two on Alice's side and two on Bob's side, see figure~\ref{fig:QKDSetup}) where to each detector click a classical binary outcome is assigned. Particular care is necessary when multi-photon states are measured \cite{PhysRevA.84.020303,PhysRevLett.101.093601}. In the following, we give the definition of the QBER for the case of photon-number-resolving detectors and we refer to \cite{amirloo_quantum_2010} for the definition in the case of threshold detectors. The probability that a particular detection pattern occurs is given by
\begin{equation}
\label{eq:detpattern}
 P_{jklm}^{(i)}:=\tr\left(\Pi_{d_0^A}^{(j)}\Pi_{d_1^A}^{(k)}\Pi_{d_0^B}^{(l)}\Pi_{d_1^B}^{(m)}\rho_{AB}^{(i)}\right),
\end{equation}
where the POVM $\Pi^{(n)}$ has been defined in \eref{eq:POVMPNRD}  with a subscript  denoting the detectors given in \fref{fig:QKDSetup}. The  superscript $i$  refers to the measurement basis and $\rho_{AB}^{(i)}$  represents the state $\rho_{AB}$ rotated in the basis $i$.

A valid QKD measurement event happens when one detector on Alice's side and one on Bob's side click. The probability of this event is given by \cite{amirloo_quantum_2010}
\begin{equation}
 P_{\rm click}^{(i)}:= P_{1010}^{(i)}+ P_{0101}^{(i)}+P_{0110}^{(i)}+P_{1001}^{(i)}.
\end{equation}
The probability that two outcomes do not coincide is given by \cite{amirloo_quantum_2010}
\begin{equation}
 P_{\rm err}^{(i)}:= P_{0110}^{(i)}+P_{1001}^{(i)}.
\end{equation}
Thus, the fraction of discordant bits, i.e., the quantum bit error rate  for measurement basis $i$ is \cite{amirloo_quantum_2010}
\begin{equation}
 e_{i}:=\frac{P_{\rm err}^{(i)}}{P_{\rm click}^{(i)}}.
\end{equation}
 For the case that $\rho_{AB}$ is a two-qubit state, we find that the QBER does not depend on the efficiency of the detectors, as $P_{\rm click}^{(i)}=\etaD^2$ and $P_{\rm err}^{(i)}\propto\etaD^2$.

If we assume a genuine two-qubit system\footnote{Note that the states of the DLCZ-type quantum repeaters
(see section~\ref{sec:atomicensenbles}) are only effectively two-qubit states,  when higher-order excitations of the atom-light entangled states \cite{duan_long-distance_2001}, or those of the states created through parametric down conversion \cite{minar}, are neglected.} like in the original quantum repeater proposal (see section~\ref{sec:briegel}) or the hybrid quantum repeater (see section~\ref{sec:Hybrid}), without loss of generality\footnote{As proven in \cite{Renner:2005pi,Kraus:2005kx}, it is possible to apply an appropriate local twirling operation that transforms an arbitrary two-qubit state into a Bell diagonal state, while the security of the protocol is not compromised.}, the entangled state $\rho_{AB}$ can be considered diagonal in the Bell-basis,
i.e., $\rho_{AB}=A\proj{\phi^+}+B\proj{\phi^-}+C\proj{\psi^+}+D\proj{\psi^-}$, with the probabilities $A, B, C, D$, $A+B+C+D=1$, and with the dual-rail\footnote{In this paper, by \emph{dual-rail representation} we mean that a single photon can be in a superposition of two optical modes, thus representing a single qubit. By \emph{single-rail representation} we mean that a qubit is implemented using only one single optical mode. See \cite{kok_introduction_2010} for additional details.} encoded Bell states\footnote{The ket $\ket{abcd}$ is a vector in a Hilbert space of four modes and the  values of $a$, $b$, $c$ and $d$ represent the number of excitations in the Fock basis.} $\ket{\phi^{\pm}}=(\ket{1010}\pm\ket{0101})/\sqrt{2}$ and $\ket{\psi^{\pm}}=(\ket{1001}\pm\ket{0110})/\sqrt{2}$ (we shall use the notation $\ket{\phi^{\pm}}$ and $\ket{\psi^{\pm}}$ for the Bell basis in any type of encoding throughout the paper). Then the QBER  along the directions $X$, $Y$,  and $Z$  corresponds to
\cite{Scarani:2009}
\begin{equation}
\label{eq:QBERBELL}
 \eci{X}:=B+D,\quad\quad
  \eci{Z}:=C+D,\quad\quad
\eci{Y}:=B+C.
\end{equation}
Throughout the whole paper $X$, $Y$ and $Z$ denote the three Pauli operators acting on the restricted Hilbert space of qubits.

\subsubsection{The secret key rate}
\label{sec:qber}
\newcommand{\rrawm}{$\rraw$}
\newcommand{\fkey}{f_{\textrm{key}}}
\newcommand{\rqkd}{R_{\rm QKD}}
\newcommand{\fkeym}{$\fkey$}
The figure of merit representing the performance of quantum key distribution is the \emph{secret key rate} $\rqkd$ which is the product of the \emph{raw key rate} $\rraw$ (see above) and the \emph{secret fraction} $r_{\infty}$. Throughout this paper, we will use asymptotic secret key rates. The secret fraction represents the fraction of secure bits that may be extracted from the raw key. Formally,  we have
\begin{equation}
\rqkd:=\rraw r_{\infty}=\rrep P_{\rm click}\rsift r_{\infty},
\label{eq:keyrate}
\end{equation}
where the sifting rate $\rsift$ is the fraction of measurements performed in the same basis by Alice and Bob Throughout the whole paper we will use $\rsift=1$ which represents the asymptotic bound for $\rsift$ when the measurement basis are chosen with biased probability \cite{Lo2005}.
We point out that both $\rrep$ and $r_{\infty}$ are functions of the explicit repeater protocol and the involved experimental parameters, as we will discuss in detail later.  Our aim is to  maximize the overall secret key rate $\rqkd$. There will be a trade-off between $\rrep$ and $r_{\infty}$, as the secret key fraction $r_{\infty}$ is an increasing function of the final fidelity, while the repeater rate $\rrep$ typically decreases with increasing final fidelity.

Note that even though for the considered protocol we find upper bounds on the secret key rate, an improved model (e.g.\ including distillation in later nesting levels or multiplexing\cite{Collins2007}) could lead to improved key rates. 


The secret fraction represents the fraction of secure bits over the total number of measured bits. We adopt the \emph{composable security definition} discussed in \cite{BenOr:2005fk,renner05, muller2009composability}. Here, composable means that  the secret key  can be used in successive tasks without compromising its security. In the following we calculate secret key rates using the state produced by the quantum repeater protocol.

In the present work, we consider only two QKD protocols, namely the BB84-protocol and the six-state protocol, for which collective and coherent attacks are equivalent \cite{Renner:2005pi, Kraus:2005kx} in the limit of a large number of exchanged signals. The unique parameter entering the formula of the secret fraction is the quantum bit error rate (QBER).

In the BB84-protocol only two of the three Pauli matrices are measured. We adopt the asymmetric protocol where the measurement operators are chosen with different probabilities \cite{Lo2005}, because this leads to higher key rates. We call $X$ the basis used for extracting a key, i.e., the basis that will be chosen with a probability of almost one in the measurement process,  while $Z$ is the basis used for the estimation of the QBER. Thus, in the asymptotic limit,  we have $\rsift=1$. The formula for the secret fraction is
\cite{Scarani:2009}
\begin{equation}
\label{eq:r:BB84}
 r_{\infty}^{\rm BB84}:=1-h(e_{Z})-h(e_{X}),
\end{equation}
with $h(p):=-p\log_2 p-(1-p)\log_2(1-p)$ being the binary entropy. This formula is an upper bound on the secret fraction, which is only achievable for ideal implementations of the protocol;
any realistic, experimental imperfection will decrease this secret key rate.

In the six-state protocol we use all three Pauli matrices. We call $X$ the basis used for extracting a key, which will be chosen with a probability of almost one, and  both $Y$  and $Z$ are the bases used for parameter estimation.  In this case, the formula for the secret fraction is given by \cite{Scarani:2009,Renner2008}\footnote{Note that the formula for the six-state protocol is independent of the choice of basis, when we assume the state of Alice and Bob $\rho_{AB}$ to be Bell diagonal. Then the secret fraction reduces to $r_{\infty}^{6S}=1-S(\rho_E)$ with $S(\rho)$ the von Neumann entropy and $\rho_E$ is the eavesdropper's state.}
\begin{align}
\label{eq:r:6S}
r^{\rm 6S}_{\infty}:=&1-e_{Z}h\left(\frac{1+(e_{X}-e_{Y})/e_{Z}}{2}\right)\nonumber\\
&-(1-e_{Z})h\left(\frac{1-(e_X+e_Y+e_Z)/2}{1-e_Z}\right)-h(e_{Z}).
\end{align}

\subsection{Methods}
The secret key rate represents the central figure of merit for  our investigations. We study the BB84-protocol, because it is most easily implementable and can also be used for protocols, where $\rho_{AB}$ is not a two-qubit state, with help of the squashing model \cite{PhysRevA.84.020303,PhysRevLett.101.093601}. Throughout the paper, we also report on results  of the six-state protocol if applicable.  We evaluate \eref{eq:keyrate} exactly,   except for the quantum repeater based on atomic ensembles where we truncate the states and cut off the higher excitations at some maximal number (see footnote~\ref{foot:dlczcalc} for  the details). For the  maximization of the secret key rate,
we have
used the numerical functions provided by Mathematica \cite{mathematica8}.

\section{The original quantum repeater}

\label{sec:briegel}
In this section, we consider a general class of quantum repeaters in the spirit of the original proposal by Briegel \etal\cite{briegel_quantum_1998}. We will analyze the requirements for the experimental parameters such that the quantum repeater is useful in conjunction with QKD. The model we consider in this section is applicable whenever two-qubit entanglement is distributed by using qubits encoded into single photons. This is the case, for instance, for quantum repeaters based on ion traps or Rydberg-blockade gates. We emphasize that we do not aim to capture all peculiarities of a specific set-up.  Instead, our intention is to present a fairly general analysis that can give an idea of the order of magnitude,  which has to be achieved for the relevant  experimental parameters. The error-model we consider is the one used in \cite{briegel_quantum_1998}.

\subsection{The set-up}

\subsubsection*{Elementary entanglement creation}
The probability that two adjacent repeater stations (separated by distance $L_0$) share an entangled pair is given by
\begin{equation}
\label{eq:briegP0}
 P_{0}:=\etaT{L_0},
\end{equation}
where $\etaT{\ell}$,  as defined in \eref{eq:etaT}, is the probability that a photon is not absorbed
 during the channel transmission.
In a specific protocol, $P_0$ may contain an additional multiplicative factor such as the probability that entanglement is heralded or also a source efficiency.
\newcommand{\bells}[1]{\psi_{#1}}
We assume that the state  created over distance $L_0$ is a depolarized state of fidelity $F_0$ with respect to $\ket{\phi^+}$, i.e.,
\begin{align}
\label{eq:briegelinitstate}
 \rho_0:=&F_0\ket{\phi^+}\bra{\phi^+}&\nonumber\\
&+\frac{1-F_0}{3}\left(\ket{\psi^+}\bra{\psi^+}+\ket{\psi^-}\bra{\psi^-}+\ket{\phi^-}\bra{\phi^-}\right).
\end{align}
The fidelity $F_0$ contains the noise due to an imperfect preparation and the noise in the quantum channel. We have chosen a depolarized state, because  this corresponds to a generic  noise model and, moreover,  any two-qubit mixed quantum state can be brought  into this form using local twirling  operations \cite{bennett_mixed-state_1996}.

\subsubsection*{Imperfect gates}
For  the local qubit operations, such as the CNOT  gates, we use  a generic gate model with depolarizing noise,  as considered in \cite{briegel_quantum_1998}. Thus,  we assume that a noisy gate $O_{BC}$  acting upon two qubits $B$ and $C$ can be modeled by
\begin{equation}
\label{eq:p2}
 O_{BC}(\rho_{BC})=p_{G}O_{BC}^{\mathrm{ideal}}(\rho_{BC})+\frac{1-p_{G}}{4}\one_{BC},
\end{equation}
where $O_{BC}^{\mathrm{ideal}}$ is the ideal gate  operation and $p_G$ describes the gate quality.
Note that, in general,  the noisy gates realized in an experiment do not necessarily have this form, however,  such a noise model is useful for having an indication  as to how good the corresponding gates must be. Other noise models  could be analogously incorporated into our analysis. Further, we assume that one-qubit gates are perfect.

\subsubsection*{Entanglement distillation}
We consider entanglement distillation only before the first entanglement  swapping steps, right after the initial pair distributions over $L_0$. We employ the Deutsch \etal protocol \cite{deutsch1996quantum} which  indeed has some advantages, as shown in the analysis of \cite{Bratzik2012}. In App.~\ref{sec:appDistill}, we review this protocol and we also present the corresponding formulas in the presence of imperfections. We point out that when starting with two copies of depolarized states, the distillation protocol will generate an output state which is no longer a depolarized state, but instead a generic Bell diagonal state. Distillation requires two-qubit gates, which we describe using \eref{eq:p2}. 

\subsubsection*{Entanglement swapping}
 The entanglement connections are performed through entanglement swapping by implementing a (noisy) Bell measurement on the photons stored in two  local quantum memories. We consider a Bell measurement  that is deterministic in the ideal case. It is implemented using a two-qubit gate with gate quality $p_G$ (see \eref{eq:p2}). Analogous to the case of distillation, starting with two depolarized states, at the end of the noisy Bell measurement, we will obtain generic Bell diagonal states. Also in this case, it turns out that a successive depolarization decreases the secret key rate and this step is therefore not performed in our scheme.

\subsection{\label{subsec:ImperfG}Performance in the presence of imperfections}
The secret key rate \eref{eq:keyrate} represents our central object of study, as it characterizes the performance of a QKD protocol. It can be written explicitly as a function of the relevant parameters,

\begin{eqnarray}
\label{eq:rqkd:briegel}
 \rqkd^{\rm O}=&&\rrep(L_0, N, k, F_0, p_G, \etaD)P_{\rm click}(\etaD)\nonumber\\
&&\times\rsift r_{\infty}(N, k, F_0, p_G),
\end{eqnarray}

where $\rrep$ is given by \eref{eq:zn:approx} when $\etaD=1$ (because then $P_{\rm ES}=1$) or by \eref{eq:avgngen} if $\etaD<1$\footnote{The supposed link between the effect of imperfect detectors and the determinism of the entanglement swapping here assumes the following. Any incomplete detection patterns that occur in the Bell measurements due to imperfect detectors are considered as inconclusive results and will be discarded. Conversely, with perfect detectors, we assume that we always have complete patterns and thus the Bell state discrimination becomes complete too. Note that this kind of reasoning directly applies to Bell measurements in dual-rail encoding, where the conclusive output patterns always have the same fixed total number for every Bell state (namely two photons leading to two-fold detection events), and so any loss of photons will result in patterns considered inconclusive. In single-rail encoding, the situation is more complicated and patterns considered conclusive may be the result of an
imperfect detection.}.
The probability that the QKD measurement is successful is given by $P_{\rm click}=\etaD^2$ and the secret fraction $r_{\infty}$ is given by either \eref{eq:r:BB84} or \eref{eq:r:6S}, depending on the type of QKD protocol. For the asymmetric BB84-protocol,  we have $\rsift=1$ (see \sref{sec:QKD}). The superscript ${\rm O}$  refers to the original quantum repeater proposal  as considered in this section.
 In order to have a non-zero secret key rate, it is then necessary that the repeater rate, the probability
for a valid QKD measurement event, and the secret fraction
are each non-zero too. As typically $\rrep > 0$, $\rsift>0$ and $P_{\rm click} > 0$, for $\rqkd >0$,
it is sufficient to have a non-zero secret fraction, $r_{\infty}>0$. The value of the secret fraction does not depend on the distance,  and therefore some properties of this protocol are distance-invariant.

\paragraph*{Minimally required parameters}
In this paragraph, we will discuss the minimal requirements that are necessary to be able to extract a secret key, i.e., we will specify the  parameter region where the secret fraction is non-zero. From the discussion in the previous paragraph, we know that this region does not depend on the total distance, but only on the initial fidelity $F_0$, the gate quality  $p_G$, the number of segments $2^N$, and the maximal number  of distillation rounds $k$. Moreover, note that even if the secret fraction is not zero, the total secret key rate can be very low  (see below).

For calculating the minimally required parameters, we start with the initial state in \eref{eq:briegelinitstate}, we distill it $k$ times (see the formulas in App.~\ref{sec:appDistill}), and then we swap the distilled state $2^N-1$ times ( see the formulas in \ref{sec:brig:es}). At the end, a generic Bell diagonal state is obtained. Using \eref{eq:QBERBELL} one can  then calculate the QBER, which is sufficient to calculate the secret fraction.

\Tref{tab:minFid} and \tref{tab:minpg} show the minimally required values for  $F_0$ and  $p_G$ for different  maximal nesting levels $N$  (i.e., different numbers of segments $2^N$) and different  numbers of rounds of distillation $k$. Throughout  these tables, we can see that for the six-state protocol, the minimal fidelity and the minimal gate quality $p_G$ are lower than for the BB84-protocol. 
Our results confirm the intuition that the larger the number of distillation rounds, the smaller the affordable initial fidelity can be (at the cost of  needing higher gate qualities).
%
\begin{table}
 \begin{tabular}{@{}l*{15}{l}}
\hline\hline \backslashbox{N}{k}
&\centre{2}{0}&\centre{2}{1}&\centre{2}{2}&\centre{2}{3}\\
\crule{9}\\
\hline &BB84&6S&BB84&6S&BB84&6S&BB84&6S\\
\hline
 0 & 0.835 & 0.810 &0.733&0.728& 0.671 & 0.669 &0.620&  0.614\\
 1 & 0.912 &0.898 & 0.821& 0.818&0.742& 0.740 & 0.669& 0.664\\
 2 & 0.955& 0.947 & 0.885 & 0.884 &0.801&0.800  & 0.713& 0.709 \\
3 & 0.977 &0.973 & 0.929 & 0.928 & 0.849 & 0.848 & 0.752& 0.749\\
4 & 0.988 & 0.987& 0.957& 0.957& 0.887&  0.887&0.788& 0.785\\
5&0.994& 0.993&0.975&  0.975& 0.917& 0.917&0.819&0.818\\
6&0.997&0.997& 0.985& 0.985&0.939&0.939&0.847& 0.846\\
7&0.999&0.998&0.992&0.992&0.956&0.956&0.872&0.870\\
\hline\hline
\end{tabular}
\caption{\label{tab:minFid}Minimal initial fidelity $F_0$ ($p_G$ is fixed to one) for extracting a secret key with maximal nesting level $N$ and number of distillation rounds $k$ for the BB84- and six-state protocols.}
\end{table}

\begin{table}
 \begin{tabular}{@{}l*{15}{l}}
\hline\hline \backslashbox{N}{k}
&\centre{2}{0}&\centre{2}{1}&\centre{2}{2}&\centre{2}{3}\\
\crule{9}\\
\hline &BB84&6S&BB84&6S&BB84&6S&BB84&6S\\
\hline
0 & - &-&0.800&0.773 & 0.869&0.860 &0.891&0.884\\
1 & 0.780&0.748 &0.922&0.910& 0.942 &0.937 & 0.947& 0.942   \\
2 & 0.920 &0.908 & 0.965& 0.960 &0.973&0.970 & 0.974& 0.972 \\
3  & 0.965& 0.959 & 0.984& 0.981 & 0.987& 0.986& 0.987& 0.986\\
4&0.984& 0.981&0.992&0.991&0.994&0.993&0.994&0.993\\
5&0.992&0.991&0.996&0.995&0.997&0.997&0.997&0.997 \\
6&0.996& 0.995&0.998&0.998&0.999&0.998&0.999&0.998\\
7&0.998&0.998&0.999&0.999&0.999&0.999&0.999&0.999\\
\hline\hline
\end{tabular}
\caption{\label{tab:minpg}Minimal $p_{G}$ ($F_0$ is fixed to one) for extracting a secret key with maximal nesting level $N$ and number of distillation rounds $k$ for the BB84- and six-state protocols.}
\end{table}

\begin{figure}[h]
 \centering
 \includegraphics[width=8cm, clip]{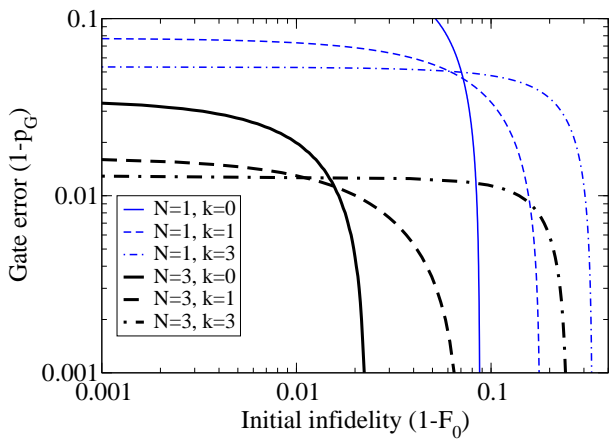}
 \caption{(Color online) Original quantum repeater and the BB84-protocol:  Maximal infidelity  $(1-F_0)$ as a function of  gate error  $(1-p_G)$ permitting to extract a secret key for various  maximal nesting levels $N$ and  numbers of distillation rounds $k$ (Parameter: $L=600$   km).}
 \label{fig:distillationNew}
\end{figure}

In \fref{fig:distillationNew}, the lines represent the values of the  initial infidelity and  the  gate error for a specific $\nl$  that allow for extracting a secret key. As shown in \fref{fig:distillationNew},  any lower initial fidelity requires a correspondingly higher gate quality and vice versa. Note that above the lines in \fref{fig:distillationNew} it is not possible to extract a secret key.


\paragraph*{The secret key rate}
In this section, we will analyze the  influence of the imperfections on the secret key rate, see \eref{eq:rqkd:briegel}.

\begin{figure}[h]
 \centering
 \includegraphics[width=8cm, clip]{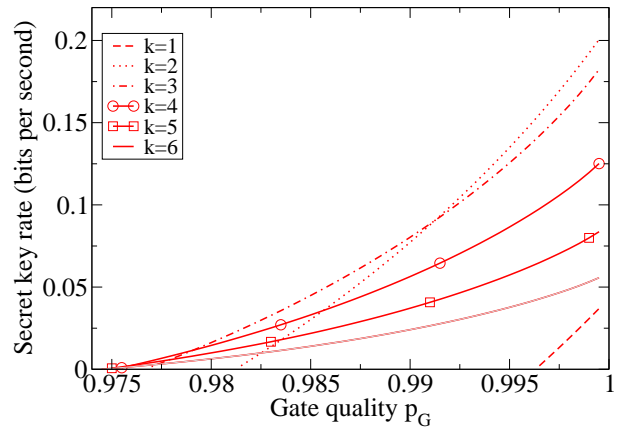}
 \caption{(Color online) Original quantum repeater and the BB84-protocol: Secret key rate \eref{eq:rqkd:briegel}
 versus gate quality  $p_G$ for different rounds of distillation $k$. The case $k=0$ leads to a vanishing secret key rate. (Parameters: $F_0=0.9$, $N=2$, $L=600$ km)}
 \label{fig:rsecvspg}
\end{figure}

In \fref{fig:rsecvspg} we  illustrate the effect of gate imperfections on the secret key rate for different  numbers of rounds of distillation and for a fixed  distance, initial fidelity, and maximal number of nesting levels. Throughout this whole section, we use $\beta=2$ in \eref{eq:fundtime} for the fundamental time, which corresponds to the case where a source is placed at one side of  an elementary segment (see \fref{fig:comm_time}). The optimal number of distillation rounds decreases as $p_G$ increases. We see from the figure that $k=2$ is optimal when $p_G=1$. This is due to the fact that from $k=1$ to $k=2$, the raw key rate decreases by $40\%$, but the secret fraction increases by $850\%$. However, from $k=2$ to $k=3$, the raw key rate decreases once again by $40\%$, but now the secret fraction increases only by $141\%$.  In this case, the net gain is smaller than 1 and therefore three rounds of distillation do not  help to increase the secret key rate compared to the
case of two rounds. In other words, what  is lost in terms of success probability  when having three probabilistic distillation rounds is not  added to the secret fraction. For  a decreasing $p_G$, more rounds of distillation become optimal.
The reason is that when the gates become worse, additional rounds of distillation permit to increase the secret key rate sufficiently much to compensate the decrease of $\rrep$.
\begin{figure}[h]
 \centering
 \includegraphics[width=8cm, clip]{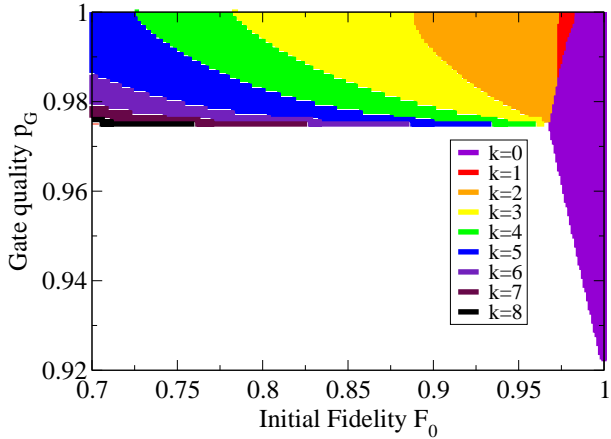}
 \caption{(Color online) Original quantum repeater  and the BB84-protocol:
Number of distillation rounds $k$ that  maximizes the secret key rate as a function of gate
quality $p_G$ and initial fidelity $F_0$. In the white area, it is  no longer possible to extract a secret key. (Parameters: $N=2$, $L=600$ km)}
 \label{fig:simulateprotocol}
\end{figure}

In \fref{fig:simulateprotocol} we show the optimal number of rounds of distillation $k$ as a function of the imperfections of the gates and the initial fidelity. It turns out that when the experimental parameters are good enough, then distillation is not necessary at all.

\begin{figure}[h]
 \centering
 \includegraphics[width=8cm, clip]{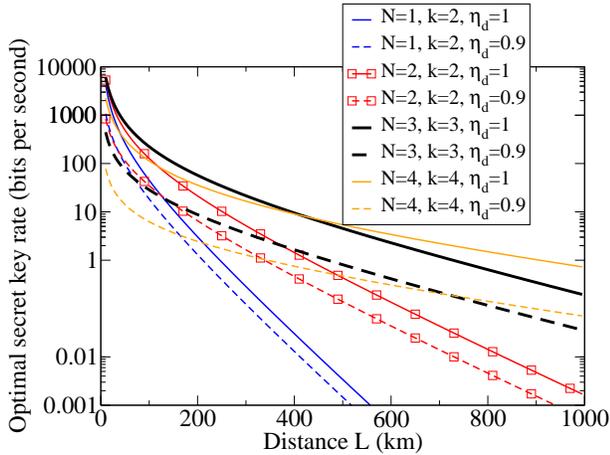}
 \caption{(Color online) Original quantum repeater and the BB84-protocol: Optimal secret key rate \eref{eq:rqkd:briegel}
 versus distance for different nesting levels, with and without perfect detectors. For each  maximal nesting level $N$, we have chosen the optimal number of distillation rounds $k$.  A nesting level $N\geq5$  no longer permits to  obtain a non-zero secret key rate. (Parameters: $F_0=0.9$ and $p_G=0.995$.)}
 \label{fig:scrvsL}
\end{figure}

 Let us now investigate the secret key rate \eref{eq:rqkd:briegel} as a function of the distance $L$ between Alice and Bob. In \fref{fig:scrvsL} the secret key rate for the optimal number of distillation rounds versus the distance for various nesting levels is shown, for a fixed initial fidelity and gate quality.  These curves should be interpreted as  upper bounds; when additional imperfections are included, the secret key rate will  further decrease. We see that for a distance of more than $400$ km, the value $N=4$ (which corresponds to 16 segments) is optimal. Note that with the initial fidelity and gate quality  assumed here, it is no longer possible to extract a secret key for $N=5$.

In many implementations, detectors are far from being perfect. The  general expression of the raw key rate  including detector efficiencies  $\etaD$ becomes
\begin{equation}
 \rraw=\frac{1}{T_0}\rsift\left(\frac{2}{3}\right)^{N+k}\etaD^{2 (k+N+1)} P_0\prod_{i=1}^{k}\pdis[i],
\label{eq:rrawES}
\end{equation}
 using \eref{eq:keyrate} with the repeater rate $R_{\rm REP}$ given by \eref{eq:avgngen}.
The term $\etaD^{2 k}$ arises from the two-fold  detections for the distillation, and similarly, $\etaD^{2N}$  comes from  the entanglement swapping and $\etaD^2$ from the QKD measurements.

In \fref{fig:scrvsL} we observe that even if detectors are imperfect, it is advantageous to do the same  number of rounds of distillation as for the perfect case. This is due to the fact that the initial fidelity is so low that even with a lower success probability, the gain in the secret fraction produces a net gain  greater than 1.

 For realistic detectors, the dark count probability is much smaller than their efficiency. We show in App.~\ref{app:general-type} that, provided that the dark count probability is smaller than $10^{-5}$, dark counts can be neglected. This  indeed applies to most modern detectors \cite{eisaman2011invited}.

\section{\label{sec:Hybrid}The hybrid quantum repeater}
In this section, we will investigate the so-called hybrid quantum repeater (HQR) introduced by van Loock \etal \cite{van_loock_hybrid_2006} and Ladd \etal \cite{ladd_hybrid_2006}. In this scheme, the resulting entangled pairs are discrete atomic qubits, but the probe system (also called \textit{qubus}) that mediates the two-qubit entangling interaction is an optical mode in a coherent state. The scheme does not only  employ atoms and light at the same time, but it also uses both discrete and continuous quantum variables; hence the name hybrid. The entangled pair is conditionally prepared by suitably measuring the probe state after it has interacted with two atomic qubits located in the two spatially separated cavities at two neighboring repeater stations. Below we shall consider a HQR where the detection is based on an unambiguous state discrimination (USD) scheme \cite{van_loock_quantum_2008, azuma}. In this case, arbitrarily high fidelities can be achieved at the expense of low
probabilities of success.

\subsection{The set-up}

\subsubsection*{Elementary entanglement creation}

\begin{figure}[h]
\centering
\includegraphics[width=8cm, clip]{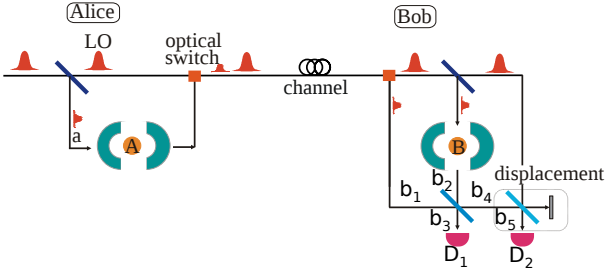}
\caption{(Color online) Schematic diagram for the entanglement generation by means of  a USD measurement following \cite{azuma}. The two quantum memories $A$ and $B$ are  separated by a distance $L_0$. The part on the left side (an intermediate Alice) prepares a pulse in a coherent state $\ket{\alpha}_a$ (the subscript refers to the corresponding spatial mode). This pulse first interacts with her qubit $A$ and is then sent to  the right side together with the local oscillator pulse (LO). The part on the right side  (an intermediate Bob) receives the state $\ket{\sqrt{\eta_t}\alpha}_{b_1}$ and produces from the LO through beam splitting a second probe pulse $\ket{\sqrt{\eta_t}\alpha}_{b_2}$ which interacts with his qubit $B$. He further applies a 50:50 beam splitter to the pulses in modes $b_1$ and $b_2$, and a displacement $D(-\sqrt{2\eta_t}\alpha\cos{\theta/2})=e^{-\sqrt{2\eta_{t}}\alpha\cos{\theta/2}(a^{\dagger}-a)}$ to the pulse in mode $b_4$. The entangled state is conditionally
generated depending on the results of detectors $D_1$ and $D_2$. The  fiber attenuation $\etaT{L_0}$ has been defined in \eref{eq:etaT}.}
\label{fig:hqr}
\end{figure}

Entanglement is shared between two electronic spins (such as $\Lambda$ systems effectively acting as two-level systems) in two distant cavities (separated by $L_0$). The entanglement distribution occurs through the interaction of the coherent-state pulse with both atomic systems. The coherent-state pulse and the cavity are in resonance, but they are detuned from the transition between the ground state and the excited state of the two-level system. This interaction can then be described by the Jaynes-Cummings interaction Hamiltonian in the limit of large detuning, $H_{int}=\hbar\chi Z a^{\dagger}a$, where $\chi$ is the light-atom coupling strength, $a$ ($a^{\dagger}$) is the annihilation (creation) operator of the electromagnetic field mode, and $Z=\ket{0}\bra{0}-\ket{1}\bra{1}$ is the $Z$ operator for a two-level atom (throughout this section, $\ket{0}$ and $\ket{1}$ refer to the two $Z$ Pauli eigenstates of the effective two-level matter system and not to the optical vacuum and
one-photon Fock states). After the interaction of the qubus in state $\ket{\alpha}$ with the first atomic state, which is initially prepared in a superposition, the output state is $U_{int}\left[\ket{\alpha}(\ket{0}+\ket{1})/\sqrt{2}\right]=(\ket{\alpha e^{-i\theta/2}}\ket{0}+\ket{\alpha e^{i\theta/2}}\ket{1})/\sqrt{2}$, with $\theta=2\chi t$ an effective light-matter interaction time inside the cavity. The qubus probe pulse is then sent through the lossy  fiber channel and interacts with the second atomic qubit also prepared in a superposition. Here we consider the protocol of \cite{azuma}, where linear optical elements and photon detectors are used for the unambiguous discrimination of the phase-rotated coherent states. Different from \cite{azuma},  however, we use imperfect photon-number-resolving detectors (PNRD), as described by \eref{eq:POVMPNRD}, instead of threshold detectors. By performing such a USD measurement on the probe state, as illustrated in \fref{fig:hqr}, the following entangled state
can be conditionally prepared,
\begin{equation}
\rho_0:=F_0\ket{\phi^+}\bra{\phi^+}+(1-F_0)\ket{\phi^-}\bra{\phi^-},
\label{initial.state}
\end{equation}
where we find $F_0=[1+e^{-2(1+\eta_t(1-2\eta_d))\alpha^2\sin^2(\theta/2)}]/2$  for $\alpha$ real, $\eta_t(L_0)$ is the  channel transmission given in \eref{eq:etaT}, and $\eta_d$ is the detection efficiency (see section~\ref{subsec:parameters}). Our derivation of the fidelity $F_0$ can be found in App.~\ref{sec:app:gen}. Note that the form of this state is different from the state considered in section~\ref{sec:briegel}. It is a mixture of only two Bell states, since the two other (bit flipped) Bell states are filtered out through the USD measurement. The remaining mixedness is due to a phase flip induced by the coupling of the qubus mode with the lossy  fiber environment.
 We find the optimal probability of success to generate an entangled pair in state $\rho_0$ 
\begin{equation}
\label{initial.p0}
 P_0=1-(2F_{0}-1)^\frac{\eta_t\eta_d}{1+\eta_t(1-2\eta_d)},
\end{equation}
which generalizes the formula for the quantum mechanically optimal USD with perfect detectors, as given in \cite{van_loock_quantum_2008}, to the case of imperfect, photon-number-resolving detectors. We explain our derivation of \eref{initial.p0} in App.~\ref{sec:app:gen}\footnote{One may also measure the qubus using homodyne detection \cite{van_loock_hybrid_2006}. However, for this scheme, final fidelities would be limited to $F_0 <0.8$ for $L_0=10$ km \cite{van_loock_hybrid_2006}, whereas by using unambiguous state discrimination, we can tune the parameters for any distance $L_0$, such that the
fidelity $F_0$ can be chosen freely and, in particular, made arbitrarily close to unity at the expense of the success probability dropping close to zero \cite{van_loock_quantum_2008}.}.

\subsubsection*{\label{sec:ESHybrid}Entanglement swapping}
A two-qubit gate is essential to perform entanglement swapping and entanglement distillation. In the HQR a controlled-Z (CZ) gate operation can be achieved by using dispersive interactions of another coherent-state probe with the two input qubits of the gate. This is similar to the initial entanglement distribution, but this time without any final measurement on the qubus \cite{van_loock_gate_2008}. Controlled rotations and uncontrolled displacements of the qubus are the essence of this scheme. The controlled rotations are realized through the same dispersive interaction as explained above. In an ideal scheme, after a sequence of controlled rotations and displacements on the qubus, the qubus mode will automatically disentangle from the two qubits and the only effect will be a sign flip on the $|11\rangle$ component of the input two-qubit state (up to single-qubit rotations), corresponding to a CZ gate operation. Thus, this gate implementation can be
characterized as measurement-free and deterministic. Using this gate, one can then perform a fully deterministic Bell measurement (i.e., one is able to distinguish between all four Bell states), and consequently, the swapping occurs deterministically (i.e., $P_{ES}\equiv1$).

In a more realistic approach, local losses will cause errors in these gates. Following \cite{louis}, after dissipation, we may consider the more general, noisy two-qubit operation $O_{BC}$ acting upon qubits $B$ and $C$,
\begin{align}
\label{gateerror}
O_{BC}(\rho_{BC})=&O_{BC}^{ideal}\left(p^2_c(x)\rho_{BC}+\right.\\ \nonumber
&\left.p_c(x)(1-p_c(x))(Z^B\rho_{BC}Z^B+Z^C\rho_{BC} Z^C)\right. \\ \nonumber
&\left. +(1-p_c(x))^2 Z^B Z^C\rho_{BC} Z^C Z^B\right),
\end{align}
where
\begin{equation}
p_c(x):=\frac{1+e^{-x/2}}{2}
\end{equation}
is the probability for each qubit to not suffer a $Z$ error, and $x:=\pi\frac{1-p_G^2}{\sqrt{p_G}(1+p_G)}$; here $p_G$ is the local transmission parameter that incorporates photon losses in the local gates.\footnote{Note that this error model is considering a CZ gate operation. For a CNOT gate, $Z$ errors can be transformed into $X$ errors.}  We derive  explicit formulas for entanglement swapping including imperfect two-qubit gates   in App.~\ref{sec:app:swap}.

\subsubsection*{Entanglement distillation}
For the distillation, the same two-qubit operation as described above in \eref{gateerror} can be used. It is then interesting to notice that if we start with a state given in \eref{initial.state}, after one round of imperfect distillation, the resulting state is a generic Bell diagonal state. The effect of gate errors in the distillation step is derived in App.~\ref{sec:app:dist}.\footnote{\label{footnote9} Note that we assume perfect qubit measurements for the distillation and the swapping, but imperfect two-qubit gates. In principle, these qubit measurements can be done using a local qubus and homodyne measurement \cite{van_loock_quantum_2008}. In this case, losses in the qubit measurement can be absorbed into losses of the gates. On the other hand, if we consider imperfect detectors for the qubit measurement then entanglement swapping will succeed with probability given by \eref{eq:briegpes}.  }

\subsection{Performance in the presence of imperfections}

In the following, we will only consider the BB84-protocol, because it is experimentally less demanding and also, because we found in our simulations that the six-state protocol produces almost the same secret key rates, due to the symmetry of the state in \eref{initial.state}. The secret key rate per second for the hybrid quantum repeater can be written as a function of the relevant parameters:

\begin{eqnarray}
R_{\rm QKD}^{\rm H}=&&R_{\rm REP}^{\rm det}(L_0,N, k,  F_0, p_G, \etaD)\nonumber\\
&&\times\rsift r_{\infty}^{\rm BB84}(L_0, N, k,  F_0, p_G),
\label{KeyH}
\end{eqnarray}
where $R_{\rm REP}^{\rm det}$ is the repeater pair-creation rate for deterministic swapping \eref{eq:zn:approx} described in section~\ref{subsec:gen} and  $r_{\infty}^{\rm BB84}$ is the secret fraction for the BB84-protocol \eref{eq:r:BB84}. For the asymmetric BB84-protocol, we have $\rsift=1$ (see \sref{sec:QKD}). The superscript ${\rm H}$ stands for hybrid quantum repeater. Note that the fundamental time is $T_0=\frac{2 L_0}{c}$, as the qubus is sent from Alice to Bob and then classical communication in the other direction is used (see section~\ref{subsec:gen} and \fref{fig:comm_time}). Further notice that the final projective qubit measurements which are necessary for the QKD protocol are assumed to be perfect. Thus, the secret key rate presented here represents an upper bound and, depending on the particular set-up adopted for these measurements, it should be multiplied by the square of the detector efficiency.

\newcommand{\foptin}{F_0^{\rm opt}}

\begin{figure}[htb]
 \centering
 \includegraphics[width=8cm, clip]{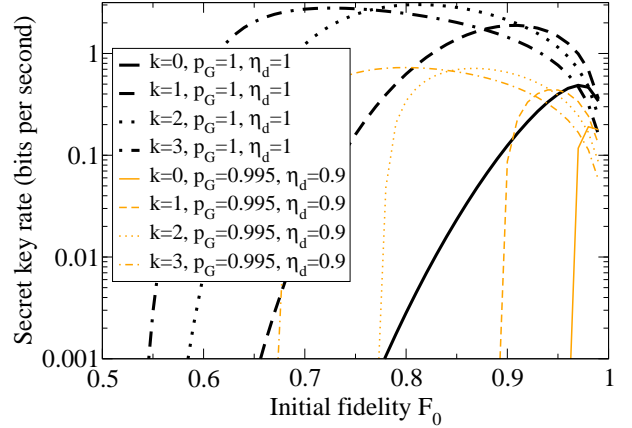}
 \caption{(Color online) Hybrid quantum repeater with perfect quantum operations ($p_G=1$) and perfect detectors ($\eta_d=1$) (black lines) compared to imperfect quantum operations ($p_G=0.995$) and imperfect detectors ($\eta_d=0.9$) (orange lines): Secret key rate per second \eref{KeyH} as a function of the initial fidelity for $2^3$ segments ($N=3$) and various rounds of distillation $k$. The distance between Alice and Bob is 600 km.}
 \label{fig:KHybridPerf}
\end{figure}

\paragraph*{The secret key rate} \Fref{fig:KHybridPerf} shows the secret key rate for $2^3$ segments ($N=3$) for various rounds of distillation. We see from the figure that for the hybrid quantum repeater the secret key rate is not a monotonic function of the initial fidelity. The reason is that increasing $F_0$ decreases $P_0$ (see \eref{initial.p0}) and vice versa. We find that the optimal initial fidelity, i.e., the fidelity where the secret key rate is maximal, increases as the maximal number of segments increases (see \Tref{tab:optF}). On the other hand, examining the optimal initial fidelity as a function of the distance, it turns out that it is almost constant for $L>100$ km. Thus, for such distances, it is neither useful nor necessary to produce higher fidelities, because these would not permit to increase the secret key rate.
%

\begin{table}
 \begin{tabular}{@{}l*{15}{l}}
\hline\hline
\backslashbox{N}{k}
&\centre{1}{0}	&\centre{1}{1}	&\centre{1}{2}	&\centre{1}{3}\\
\hline
1&0.898 &0.836	&0.765	&0.705\\
2&0.946	&0.876	&0.788	&0.715\\
3&0.972	&0.907 	&0.812	&0.726\\
4&0.986	&0.931	&0.834	&0.741\\
\hline\hline
 \end{tabular}
\caption{\label{tab:optF}Hybrid quantum repeater without imperfections ($p_G=1$ and $\eta_d=1$): Initial fidelity $F_0$ that maximizes the secret key rate in \eref{KeyH} for a given number $2^N$ of segments and $k$ rounds of distillation.}
\end{table}

We also observe that the maximum of the initial fidelity is quite broad for small $N$, and gets narrower as $N$ increases. If we now consider perfect gates and perfect detectors, we see that by fixing a certain secret key rate, we can reach this value with lower initial fidelities by performing distillation. Furthermore, by distilling the initial entanglement, we can even exceed the optimal secret key rate without distillation by one order of magnitude. However, note that distillation for $k$ rounds requires $2^k$ memories at each side. If we then assume that we choose the protocol with no distillation and perform it in parallel $2^k$ times, i.e., we use the same amount of memories
as for the scheme including distillation, the secret key rate without distillation (as shown in \fref{fig:KHybridPerf}) should be multiplied by $2^k$. As a result, the total secret key rate can then be even higher than that obtained with distillation.

\begin{figure}
 \centering
 \includegraphics[width=8cm, clip]{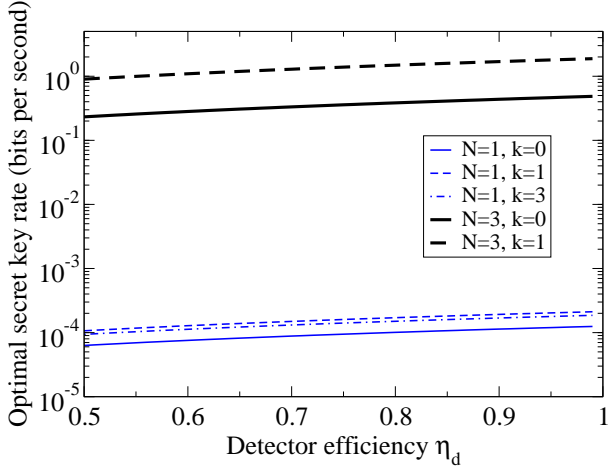}
 \caption{(Color online) Hybrid quantum repeater with perfect gates ($p_G=1$): The optimal secret key rate \eref{KeyH} for the BB84-protocol in terms of the detector efficiency $\eta_d$ for the distance $L=600$ km with various numbers of segments $2^N$ and rounds of distillation $k$.}
 \label{fig:KeyEta}
\end{figure}

Let us now assess the impact of the gate and detector imperfections on the secret key rate (orange lines) in \fref{fig:KHybridPerf}. We notice that $p_G$ has a large impact even if it is only changed by a small amount, like here from $p_G=1$ to $p_G=0.995$; the secret key rates drop by one order of magnitude. Imperfect detectors are employed in the creation of entanglement. As we see in \fref{fig:KeyEta}, imperfect detectors do not affect the secret key rate significantly. As for $N=3$ and $k=0$, improving the detector efficiency from $0.5$ to $1$ leads to a doubling of the secret key rate. We conclude that for the hybrid quantum repeater, the final secret key rates are much more sensitive to the presence of gate errors than to inefficiencies of the detectors. However, recall that in our analysis, we only take into account detector imperfections that occur during the initial USD-based entanglement distribution.
For simplicity, any measurements on the memory qubits performed in the local circuits for swapping and distillation are assumed to be perfect, whereas the corresponding two-qubit gates for swapping and distillation are modeled as imperfect quantum operations (see footnote~\ref{footnote9} for more details).

\begin{figure}[htb]
 \centering
 \includegraphics[width=8cm,clip]{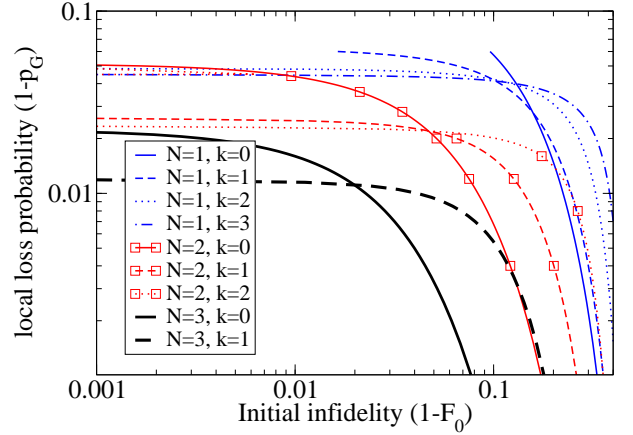}
 \caption{(Color online) Hybrid quantum repeater with distillation and imperfections:  Maximally allowed infidelity $(1-F_0)$ as a function of the local loss probability  $(1-p_G)$ for various maximal numbers of segments $2^N$ and rounds of distillation $k$ (distance: $L=600$ km).  Above the curves it is no longer possible to extract a secret key. The lines with $k=0$ correspond to entanglement swapping without distillation.}
 \label{fig:MinFidLocalT}
\end{figure}

\paragraph*{Minimally required parameters} As we have seen in the previous section, it is also worth finding the minimal parameters for $F_0$ and $p_G$, for which we can extract a secret key. \Fref{fig:MinFidLocalT} shows the initial infidelity required for extracting a secret key as a function of the local loss probability $p_G$, which was introduced in \sref{sec:ESHybrid}.
\newcommand{\tnmin}[1]{p_{G,{#1}}^{\rm min}}
We obtain also the minimal values of the local transmission probability $\tnmin{N}$ without distillation (solid lines in \fref{fig:MinFidLocalT}). If $p_G<\tnmin{N}$, then it is no longer possible to extract a secret key. As shown in \fref{fig:MinFidLocalT}, these minimal values (for which the minimal initial fidelity becomes $F_0=1$, without distillation) are $\tnmin{1}=0.853$ (not shown in the plot), $\tnmin{2}=0.948$,  $\tnmin{3}=0.977$, and $\tnmin{4}=0.989$ (not shown in the plot). When including distillation, we can extend the regime of non-zero secret key rate to smaller initial fidelities at the cost of better local transmission probabilities.
So there is a trade-off: if we can produce almost perfect Bell pairs, that is initial states with high fidelities $F_0$, we can afford larger gate errors. Conversely, if high-quality gates are available, we may operate the repeater with initial states having a lower fidelity. Note that these results and \fref{fig:MinFidLocalT} do not depend on the length of each segment in the quantum repeater, but only on the number of segments.

\begin{figure}[htb]
 \centering
 \includegraphics[width=8cm, clip]{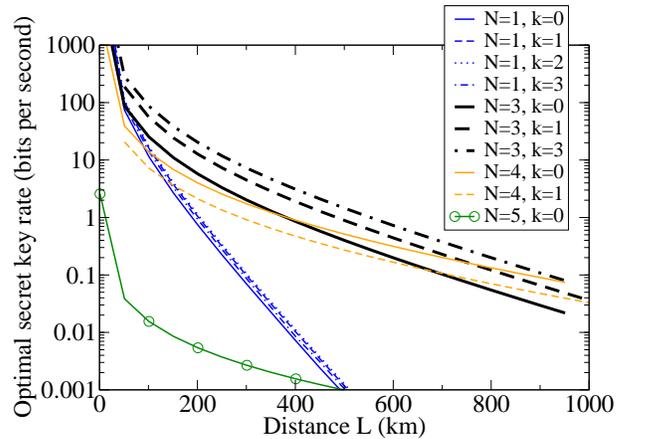}
 \caption{(Color online) Hybrid quantum repeater with imperfect quantum operations ($p_G=0.995$) and imperfect detectors ($\eta_d=0.9$): Optimal secret key rate \eref{KeyH} for the BB84-protocol as a function of the total distance $L$, for various numbers of segments $2^N$ and rounds of distillation $k$. For $N=5$, it is not possible to obtain a secret key when distillation is applied.}
 \label{fig:ImpKeyVsL}
\end{figure}

In figure~\ref{fig:ImpKeyVsL} we plotted the optimal secret key rate for a fixed local transmission probability $p_G$ and detector efficiency $\eta_d$ in terms of the total distance $L$. We varied the number of segments $2^N$ and the number of distillation rounds $k$. We observe that a high value of $k$ is not always advantageous: There exists for every $N$ an optimal $k$, for which we obtain the highest key rate. We see, for example, that for $N=1$, the optimal choice is $k=2$, whereas for $N=3$, the optimal $k$ is 3. One can also see that there are distances, where it is advantageous to double the number of segments if one wants to avoid distillation, as, for example, for $N=3$ and $N=4$ at a distance of around 750 km.

\section{Quantum repeaters based on atomic ensembles}
\label{sec:atomicensenbles}
The probably most influential proposal for a practical realization of quantum repeaters was made in \cite{duan_long-distance_2001} and it is known as the Duan-Lukin-Cirac-Zoller (DLCZ)-protocol.  These authors  suggested to use atomic ensembles as quantum memories and linear optics combined with single-photon detection for entanglement distribution, swapping, and (built-in) distillation. This proposal influenced experiments and theoretical investigations and led to improved protocols based on atomic ensembles and linear optics (see \cite{sangouard_quantum_2011} for a recent review).

To our knowledge, the most efficient scheme based on atomic ensembles and linear optics was proposed very recently by  Miná\v{r} \etal \cite{minar}. These authors suggest to use heralded qubit amplifiers \cite{ralph} to produce entanglement on demand and then to extend it using entanglement swapping based on two-photon detections. The state produced at the end of the protocol no longer contains vacuum components and therefore can be used directly for QKD. This is an improvement over the original DLCZ protocol in which the final long-distance pair is still contaminated by a fairly large vacuum term that accumulates during the imperfect storage and swapping processes.\footnote{Very recently it was shown that in the context
of QKD over continuous variables, an effective suppression of channel losses and imperfections
can also be achieved via a virtual, heralded amplification on the level of the classical
post-processing \cite{2012arXiv1205.6933F,2012arXiv1206.0936W}.   In this case, it is not even necessary to physically realize a heralded amplifier.}

In this section, we first review the protocol proposed in \cite{minar} and then we analyze the role of the parameters and the performance in relation to QKD.

\subsection{The set-up}

The protocol is organized in three logical steps. First, local entanglement is created in a repeater station, then it is distributed, and finally it is extended over the entire distance \cite{minar}.
\begin{figure}
\centering
\includegraphics[width=8cm, clip]{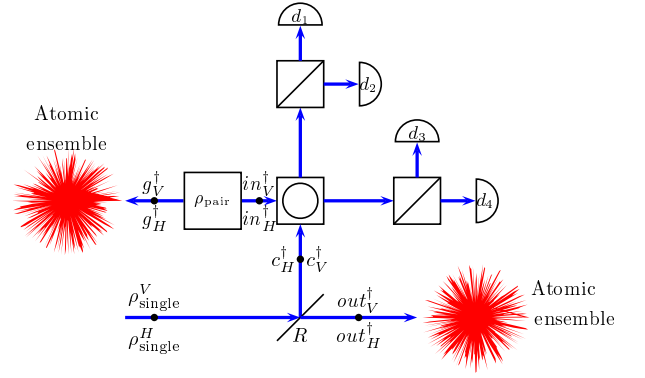}
\caption{Quantum repeater based on atomic ensembles: Set-up for creation of on-demand entanglement (see also \cite{minar}). The whole set-up is situated at one physical location. A pair source produces the state $\rho_{\rm pair}$. One part of the pair (the mode $g$) is stored in an atomic ensemble and the other part (mode $in$) goes into a linear-optics network. A single-photon source produces the states $\rho_{\rm single}^{H}$ and $\rho_{\rm single}^{V}$ which go through a beam splitter of reflectivity $R$. The output modes of the beam splitter are called $c$ and $out$. The mode $out$ is stored in a quantum memory and the mode $c$ goes into a linear-optics network which is composed of a polarizing beam splitter in the diagonal basis $\pm 45^{\circ}$ (square with a circle inside), two polarizing beam splitters in the rectilinear basis (square with a diagonal line inside), and four detectors.}
\label{fig:minarentprod}
\end{figure}

As a probabilistic entangled-pair source we consider spontaneous parametric down-conversion (SPDC) \cite{PhysRevLett.75.4337} which produces the state (see \cite{PhysRevA.61.042304} and \cite{minar})\footnote{\label{foot:dlczcalc}In our calculation, similar to \cite{minar}, we consider only those terms with $m\leq2$. The reason is that the contribution to the total trace of the first three terms is given by  $1-p^3$ and therefore for $p<0.1$ the state obtained by considering only the first three terms  differs in a negligible way from the full state.}
\newcommand{\opc}[1]{#1^{\dagger}}
\begin{equation}
\label{eq:spdc}
 \rho_{\rm pair}:=(1-p)\sum_{m=0}^{\infty} \frac{2^m p^m}{(m!)^2 (m+1)} (\opc{B})^{m}\ket{0}\bra{0}B^m,
\end{equation}
where $\opc{B}:=(\opc{g_{H}}\opc{in_{H}}+\opc{g_{V}}\opc{in_{V}})/\sqrt{2}$. The operator $g_{i}^{\dagger}$ ($in_{i}^{\dagger}$)  denotes a spatial mode with polarization given by $i=H,V$. The \emph{pump parameter} $p$  is related to the probability to have an $n$-photon pulse by $P(n)=p^n(1-p)$.

A probabilistic single-photon source with efficiency $q$ produces states of the form
\begin{equation}
 \rho_{\rm single}^{i}:=(1-q)\ket{0}\bra{0}+qa_{i}^{\dagger}\ket{0}\bra{0}a_{i},
\end{equation}
where $a^{\dagger}_{i}$($a_{i}$) is the creation (annihilation) operator of a photon with polarization $i=H, V$.

We also define by $\gamma_{\rm rep}$ the smallest repetition rate among the repetition rates of the SPDC source and the single-photon sources.

\subsubsection*{On-demand entanglement source}

The protocol that produces local entangled pairs works as follows (see \fref{fig:minarentprod} and \cite{minar} for additional details):
\begin{enumerate}
 \item The state $\rho_{\rm pair}\otimes\rho_{\rm single}^{H}\otimes\rho_{\rm single}^{V}$ is produced.
 \item The single photons, which are in the same spatial mode, are sent through a tunable beam splitter of reflectivity $R$ corresponding to the transformation $a_{i}\to \sqrt{R}~c_{i} + \sqrt{1-R}~out_{i}$.
 \item The spatial modes $in$ and $c$ are sent through a linear-optics network which is part of the heralded qubit amplifiers, and the following transformations are realized,
\begin{eqnarray*}
c_{H}\to\frac{d_3+d_4+d_2-d_1}{2},\\ c_{V}\to\frac{d_3+d_4-d_2+d_1}{2},\\
in_{H}\to\frac{d_2+d_1+d_3-d_4}{2},\\ in_{V}\to\frac{d_2+d_1-d_3+d_4}{2},
\end{eqnarray*}
where $d_1,\; d_2,\; d_3,\; d_4$ are four spatial modes, corresponding to the four detectors.
\item A  twofold coincidence detection between $d_1$ and $d_3$ (or $d_1$ and $d_4$ or $d_2$ and $d_3$ or $d_2$ and $d_4$) projects the modes $g$ and $out$ onto an entangled state. These are the heralding events that acknowledge the storage of an entangled pair in the quantum memories $out$ and $g$.The probability of a successful measurement is given by
\begin{widetext}
\begin{equation}
 P_0^{s}(p, q, R, \etaD)=4\tr\left(\Pi^{(1)}_{d_1}(\etaD)\Pi^{(0)}_{d_2}(\etaD) \Pi^{(1)}_{d_3}(\etaD)\Pi^{(0)}_{d_4}(\etaD) \rho'_{g, out, d_1, d_2, d_3, d_4}\right),
\end{equation}
\end{widetext}
 
where $\rho'_{g, out, d_1, d_2, d_3, d_4}$ is the total state obtained at the end of step (iii) and the superscript $s$ stands for source. The POVM for the detectors has been defined in \eref{eq:POVMPNRD}. The factor $4$ accounts for the fact that there are four possible twofold coincidences.
The resulting state is
\begin{widetext}
\begin{equation}
\label{eq:rho0s}
 \rho_{0}^{s}(p, q, R, \etaD)=\frac{4}{P_0^{s}} \tr_{d_1, d_2, d_3, d_4}\left(\Pi^{(1)}_{d_1}(\etaD)\Pi^{(0)}_{d_2}(\etaD) \Pi^{(1)}_{d_3}(\etaD)\Pi^{(0)}_{d_4}(\etaD) \rho'_{g, out, d_1, d_2, d_3, d_4}\right).
\end{equation}
\end{widetext}
This is the locally prepared state that will be distributed between the repeater stations.
In the ideal case with perfect detectors and perfect single-photon sources, the resulting state (after a suitable rotation) is $\rho_{0}^{s}=\ket{\phi^{+}}\bra{\phi^{+}}$ which can be obtained with probability $P_0^s=pR(1-R)$. In the realistic case, however, additional higher-order excitations are present. In \cite{minar}, the explicit form of $\rho_{0}^{s}$ and $P_{0}^{s}$ can be found for the case when $1>R\gg p$ and $1\gg 1-q$.
\end{enumerate}
Therefore, we have seen that the protocol proposed in \cite{minar} permits to turn a probabilistic entangled-pair source (SPDC in our case) into an on-demand entangled photon source. In this context \emph{on-demand} means that when a heralding event is obtained then it is known for sure that an entangled quantum state is stored in the quantum memories $out$ and $g$.

\subsubsection*{Entanglement distribution and swapping}

\begin{figure}[h!]
\centering
\includegraphics[width=.3\textwidth, clip]{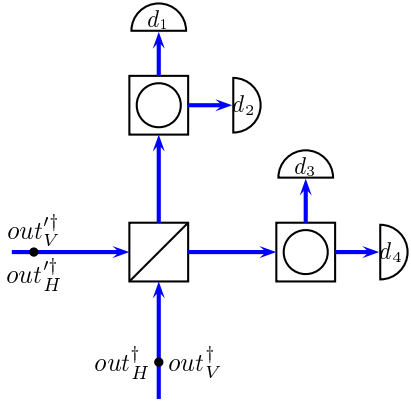}
\caption{Quantum repeater based on atomic ensembles: Set-up used for entanglement distribution (swapping) (see \cite{minar} for additional details). The modes $out$ and $out'$ are released from two quantum memories separated by distance $L_0$ (or located at the same station for the case of swapping) and sent into a linear-optics network consisting of one polarizing beam splitter in the rectilinear basis (square with diagonal line inside), two polarizing beam splitters in the diagonal basis (square with circle inside), and four detectors.}
\label{fig:minardistr}
\end{figure}
Once local entangled states are created, it is necessary to distribute the entanglement over segments of length $L_0$ and then to perform entanglement swapping. Both procedures are achieved in a similar way (see \fref{fig:minardistr}), as we shall describe in this section. Entanglement distribution is done as follows (see \fref{fig:minardistr} and \cite{minar} for additional details):
\begin{enumerate}
 \item Each of the two adjacent stations create a state of the form $\rho_{0}^{s}$. We call $g$ and $out$ the modes belonging to the first station and $g'$ and $out'$ the modes of the second station.
\item The modes $out$ and $out'$ are read out from the quantum memories and sent through an optical fiber to a central station where a linear-optics network is used in order to perform entanglement swapping. The transformations of the modes are as follows:
\begin{eqnarray*}
out_{H}\to\frac{d_3+d_4}{\sqrt{2}},\quad\ out_{V}\to\frac{d_1-d_2}{\sqrt{2}},\\
out'_{H}\to\frac{d_1+d_2}{\sqrt{2}},\quad out'_{V}\to\frac{d_3-d_4}{\sqrt{2}},
\end{eqnarray*}
where $d_1,\; d_2,\; d_3,\; d_4$ are four spatial modes.
\item
A  twofold coincidence detection between $d_1$ and $d_3$ (or $d_1$ and $d_4$ or $d_2$ and $d_3$ or $d_2$ and $d_4$) projects the modes $out$ and $out'$ onto an entangled state. The probability of this event is given by
\begin{widetext}
\begin{equation}
 P_0(p, q, R, \etaD, \eta_{\rm mtd})=4\tr\left(\Pi^{(1)}_{d_1}(\eta_{mtd})\Pi^{(0)}_{d_2}(\eta_{mtd}) \Pi^{(1)}_{d_3}(\eta_{mtd})\Pi^{(0)}_{d_4}(\eta_{mtd}) \rho'_{g, g', d_1, d_2, d_3, d_4}\right),
\end{equation}
\end{widetext}
where $\rho'_{g, g', d_1, d_2, d_3, d_4}$ is the total state obtained at the end of step (ii) and $\eta_{mtd}:=\eta_{m}\etaT{\frac{L_0}{2}}\etaD$, with $\eta_{m}$ being the probability that the quantum memory releases a photon. The factor $4$ accounts for the fact that there are four possible twofold coincidences.
The resulting state is
\begin{widetext}
\begin{equation}
 \rho_{0, g, g'}=\frac{4}{P_0} \tr_{d_1, d_2, d_3, d_4}\left(\Pi^{(1)}_{d_1}(\eta_{mtd})\Pi^{(0)}_{d_2}(\eta_{mtd}) \Pi^{(1)}_{d_3}(\eta_{mtd})\Pi^{(0)}_{d_4}(\eta_{mtd}) \rho'_{g, g', d_1, d_2, d_3, d_4}\right).
\end{equation}
\end{widetext}
\end{enumerate}
The state $\rho_{0, g, g'}$ is the entangled state shared between two adjacent stations over distance $L_{0}$. In order to perform entanglement swapping, the same steps as described above are repeated until those two stations separated by distance $L$ are finally connected. Formally, the probability that entanglement swapping is successful in the nesting level $n$ is given by
\begin{widetext}
\begin{equation}
  P_{ES}^{(n)}(p, q, R, \etaD, \eta_{\rm mtd})=4\tr\left(\Pi^{(1)}_{d_1}(\eta_{md})\Pi^{(0)}_{d_2}(\eta_{md}) \Pi^{(1)}_{d_3}(\eta_{md})\Pi^{(0)}_{d_4}(\eta_{md}) \rho'_{n-1,g, g', d_1, d_2, d_3, d_4}\right),
\end{equation}
\end{widetext}
where $\rho'_{n-1,g, g', d_1, d_2, d_3, d_4}$ is the total state resulting from steps (i) and (ii) described above in this section, and $\eta_{md}:=\eta_{m}\etaD$.

The swapped state is given by
\begin{widetext}
\begin{equation}
 \rho_{k, g, g'}=\frac{4}{P_{ES}^{(i)}} \tr_{d_1, d_2, d_3, d_4}\left(\Pi^{(1)}_{d_1}(\eta_{md})\Pi^{(0)}_{d_2}(\eta_{md}) \Pi^{(1)}_{d_3}(\eta_{md})\Pi^{(0)}_{d_4}(\eta_{md}) \rho'_{k-1,g, g', d_1, d_2, d_3, d_4}\right).
\end{equation}
\end{widetext}

%
The state $\rho_{n, g, g'}$ is the state that will be used for quantum key distribution when $n = N$. In a regime where higher-order excitations can be neglected, the state $\rho_{n, g, g'}$ is a maximally entangled Bell state. In \cite{minar} it is given the expression of the state  $\rho_{n, g, g'}$ under the same assumptions on the reflectivity $R$ and the efficiency $q$ of the single-photon sources as discussed regarding $\rho_0^s$ in \eref{eq:rho0s}.

Given the final state $\rho_{AB}:= \rho_{N, g, g'}$ it is possible to calculate $P_{\rm click}$ and the QBER, using the formalism of \sref{sec:qber} and inserting $\eta_{md}$ for the detector efficiency.

The final secret key rate then reads
\begin{widetext}
\begin{equation}
 \rqkd^{\rm AE}=\rrep(L_0, p, N, \etaD, \eta_{m},\gamma_{\rm rep}, q)P_{\rm click}(L_0, p, N, \etaD, \eta_{m}, q)\rsift r_{\infty}^{\rm BB84}(L_0, p, N, \etaD, \eta_{m}, q),
\end{equation}
\end{widetext}
where $\rrep$ is given by \eref{eq:avgngen} with $\beta=1$ for the communication time (see \fref{fig:comm_time}c). As for the QKD protocol, we consider the asymmetric BB84-protocol ($\rsift=1$, see \sref{sec:QKD}). The superscript ${\rm AE}$ stands for atomic ensembles.

Note that even though for the explicit calculations we used PNRD, the previous formulas hold for any type of measurement.
%

\subsection{Performance in the presence of imperfections}
%

As in the previous sections, we shall focus on the secret key rate. The free parameters are the pump parameter $p$ and the reflectivity of the beam splitter $R$. In all plots, we optimize these parameters in such a way that the secret key rate is maximized. As all optimizations have been done numerically, our results may not correspond to the global maximum, but only to a local maximum. In general, we observed that if we treat the secret key rate as a function of $p$ (calculated at the optimal $R$), the maximum of the secret key rate is rather narrow. On the other hand, when calculated as a function of $R$ (at the optimal $p$), this maximum is quite broad.

\begin{figure}[h]
\centering
\includegraphics[width=8cm, clip]{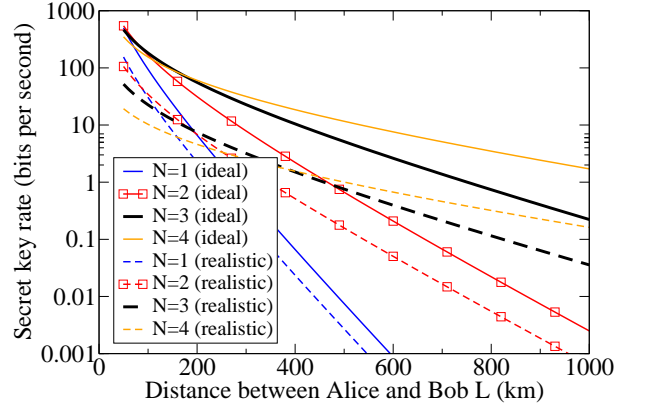}
\caption{(Color online) Quantum repeaters based on atomic ensembles: Optimal secret key rate per second versus the distance between Alice and Bob. The secret key rate has been obtained by maximizing over $p$ and $R$. Ideal set-up (solid line) with parameters $\eta_{m}=\etaD=q=1,\gamma_{rep}=\infty$. More realistic set-up (dashed line) with parameters $\eta_{m}=1$, $\etaD=0.9$, $q=0.96$, $\gamma_{rep}=50$ MHz.  }
\label{fig:optscrminar}
\end{figure}

The most favorable scenario (ideal case) is characterized by perfect detectors ($\etaD=1$), perfect quantum memories ($\eta_{m}=1$), and deterministic single-photon sources ($q=1$) which can emit photons at an arbitrarily high rate ($\gamma_{\rm rep}=\infty$). In this case, the heralded qubit amplifier is assumed to be able to create perfect Bell states and the secret fraction therefore becomes one. The only contribution to the secret key rate is then given by the repeater rate. In \fref{fig:optscrminar} the optimal secret key rate versus the distance, obtained by maximizing over $p$ and $R$, is shown (see solid lines). 

\begin{figure}[h]
\centering
\includegraphics[width=8cm, clip]{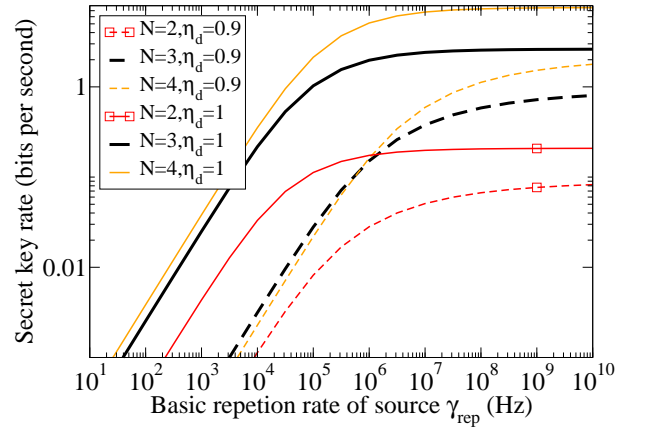}
\caption{(Color online) Quantum repeaters based on atomic ensembles: Optimal secret key rate per second versus the basic repetition rate of the source $\gamma_{\rm rep}$. The secret key rate has been obtained by maximizing over $p$ and $R$.  (Parameters: $\etaD=\eta_{m}=q=1$).}
\label{fig:scroptgamma}
\end{figure}

For the calculation of \fref{fig:optscrminar}, we have assumed that the creation of local entanglement, i.e., of state $\rho_{0}^{s}$, is so fast that we can neglect the creation time. In the case of SPDC, the repetition rate of the source is related to the pump parameter $p$ and, moreover, the single-photon sources also have finite generation rates that should be taken into account. For this purpose, we introduce the photon-pair preparation time which is given by $T_{0}^{s}=\frac{1}{\gamma_{\rm rep} P_0^{s}}$ \cite{minar}. The formula for the repeater rate in this case corresponds to \eref{eq:avgngen} with $T_0\to T_0+T_0^s$.  As shown in \fref{fig:scroptgamma}, when $\etaD=1$ the secret key rate is constant for $\gamma_{rep}>10^7$, however, for realistic detectors with $\etaD=0.9$, much higher repetition rates are required in order to reach the asymptotic value. Nowadays, SPDC sources reach a rate of about 100 MHz, whereas single-photon sources have a repetition rate of a few MHz \
\cite{eisaman2011invited}. Recently, a new  single-photon source with repetition rate of 50 MHz has been realized \cite{lee2011planar}. In the following, we will employ $\gamma_{rep}=50$ MHz.

\begin{figure}[h]
\centering
\includegraphics[width=8cm, clip]{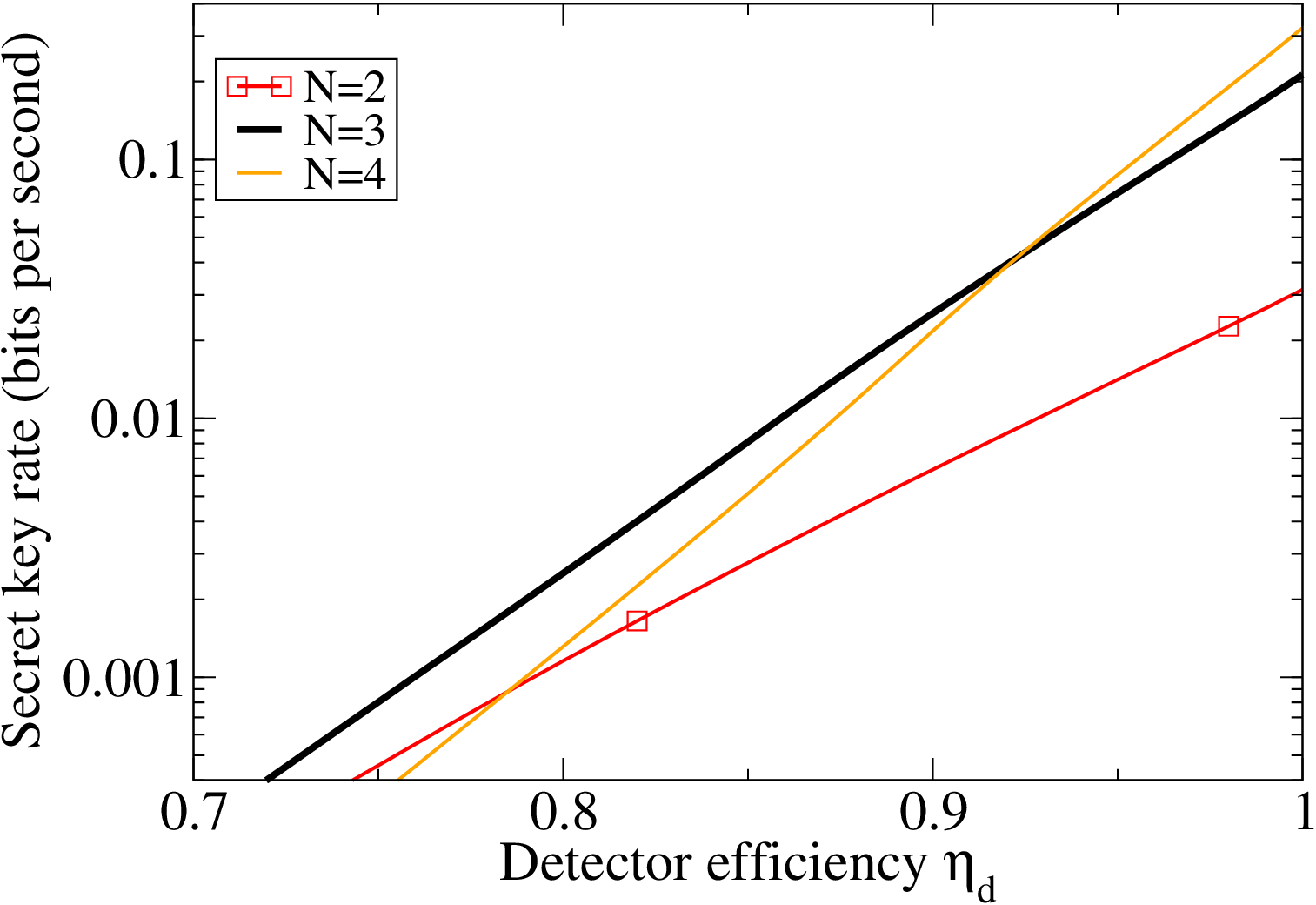}
\caption{(Color online) Quantum repeaters based on atomic ensembles: Optimal secret key rate per second versus the efficiency of the detectors $\etaD$.  The secret key rate has been obtained by maximizing over $p$ and $R$.  (Parameters: $\eta_{m}=q=1$, $\gamma_{\rm rep}=50$ MHz, $L=600$ km). }
\label{fig:scroptetaD}
\end{figure}

A consequence of imperfect detectors is that multi-photon pulses contribute to the final state. The protocol we are considering here is less robust against detector inefficiencies than the original DLCZ protocol. This is due to the fact that successful entanglement swapping is conditioned on twofold detection as compared to one-photon detection of the DLCZ protocol. However, twofold detections permit to eliminate the vacuum in the memories \cite{sangouard_quantum_2011},   thus increasing the final secret key rate. As shown in \fref{fig:scroptetaD}, the secret key rate spans four orders of magnitude as $\etaD$ increases from $0.7$ to $1$. Thus, an improvement of the detector efficiency causes  a considerable increase of the secret key rate. For example, for $N=3$, an improvement from $\etaD=0.85$ to $\etaD=0.88$ leads to a threefold increase of the secret key rate. Notice that we have considered photon detectors which are able to resolve photon numbers. Photon detectors with an efficiency as high 
as 95\% have been realized \cite{lita2008counting}.  These detectors work at the telecom bandwidth of 1556 nm and they have negligible dark counts. The drawback is that they need to operate at very low temperatures of 100 mK. The reading efficiency of the quantum memory $\etaM$ plays a similar role as the detector efficiency. In accordance to \cite{sangouard_quantum_2011}, intrinsic quantum memory efficiencies above 80\% have been realized \cite{PhysRevLett.98.190503}; however, total efficiencies where coupling losses are included are much lower.

\begin{figure}[h]
\centering
\includegraphics[width=8cm, clip]{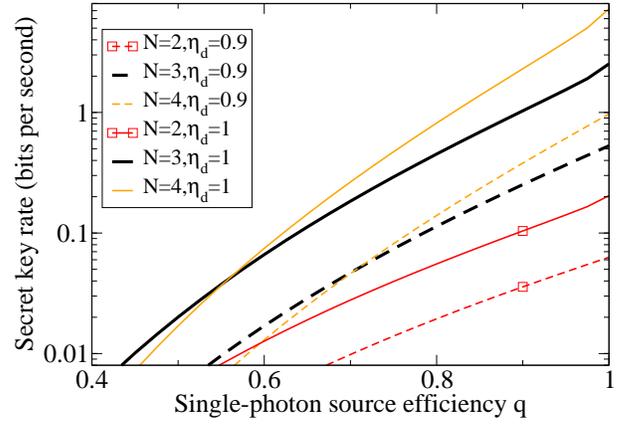}
\caption{(Color online) Quantum repeaters based on atomic ensembles: Optimal secret key rate per second versus the probability to emit a single photon. The secret key rate has been obtained by maximizing over $p$ and $R$.  (Parameters: $\eta_{m}=1, \gamma_{rep}=50$ MHz, $L=600$ km).}
\label{fig:scroptq}
\end{figure}

A single-photon source is also characterized by its efficiency, i.e., the probability $q$ to emit a photon. As shown in \fref{fig:scroptq}, we see that it is necessary to have single-photon sources with high efficiencies, in particular, when detectors are imperfect. The source proposed in \cite{lee2011planar} reaches $q=0.96$.

\begin{figure}[h]
\centering
\includegraphics[width=8cm, clip]{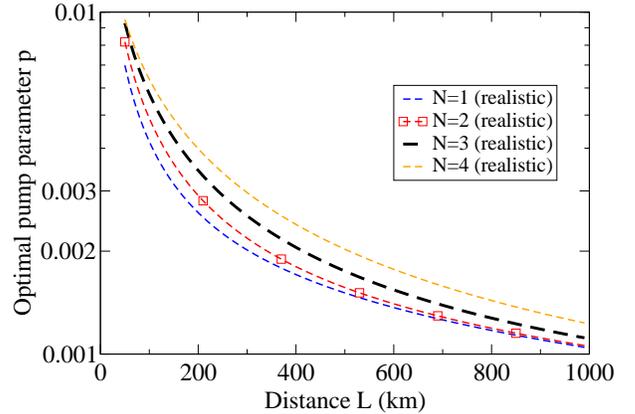}
\caption{(Color online) Quantum repeaters based on atomic ensembles: Optimal value of $p$ versus the distance between Alice and Bob. The corresponding secret key rate is shown in \fref{fig:optscrminar}.  (Parameters: $\eta_{m}=1$, $\eta_D=0.9$, $q=0.96$, $\gamma_{rep}=50$ MHz, $L=600$ km)}
\label{fig:scroptp}
\end{figure}

\begin{figure}[h]
\centering
\includegraphics[width=8cm, clip]{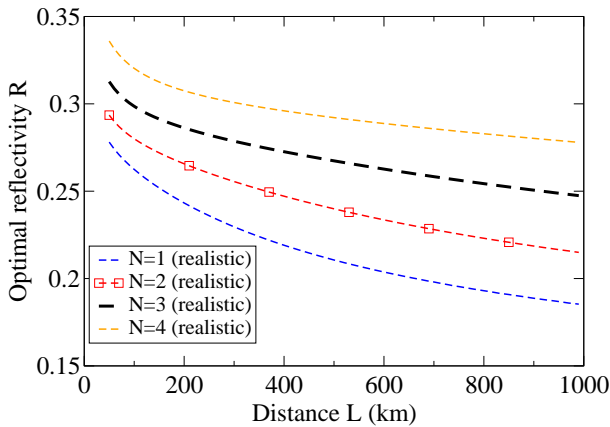}
\caption{(Color online) Quantum repeaters based on atomic ensembles: Optimal value of the reflectivity $R$  versus the distance between Alice and Bob. The corresponding secret key rate is shown in \fref{fig:optscrminar}.  (Parameters: $\eta_{m}=1$, $\eta_D=0.9$, $q=0.96$, $\gamma_{rep}=50$ MHz)}
\label{fig:scroptR}
\end{figure}

In \fref{fig:optscrminar} we show the secret key rate as a function of the distance between Alice and Bob for parameters (dashed lines) which are optimistic in the sense that they could be possibly reached in the near future. We observe that with an imperfect set-up and for $N=4$, the realistic secret key rate is by one order of magnitude smaller than the ideal value. This decrease is mainly due to finite detector efficiencies. For $N=4$, the secret key rate scales proportionally to $\etaD^2\etaD^2\etaD^{2\cdot4}\etaD^2$ (local creation, distribution, entanglement swapping, and QKD measurement). For $\etaD=0.9$, finite detector efficiencies lead to a decrease of the secret key rate by $78\%$. Regarding the optimal pump parameter $p$, we observe in \fref{fig:scroptp} that for large distances ($L>600$km) its value is about $0.15\%$. The order of magnitude of this value is in agreement with the results found in \cite{amirloo_quantum_2010} regarding the original DLCZ protocol and the BB84-protocol.
The optimal reflectivity $R$ is given in \fref{fig:scroptR}. We observe that as $N$ increases, the optimal value of $R$ has a modest increase.

\section{\label{sec:conclusions} Conclusions and Outlook }

Quantum repeaters represent nowadays the most promising and advanced approach to create long-distance entanglement. Quantum key distribution (QKD)  is a developed technology which has already reached the market. One of the main limitations of current QKD is that the two parties have a maximal separation of 150 km, due to losses in optical fibers. In this paper, we have studied long-distance QKD by using quantum repeaters.

We have studied three of the main protocols for quantum repeaters, namely, the original protocol, the hybrid quantum repeater, and a variation of the so-called DLCZ protocol. Our analysis differs from previous treatments, in which only final fidelities have been investigated, because we  maximize the main figure of merit for QKD -- the secret key rate. Such an optimization is non-trivial, since there is a trade-off between the repeater pair-generation rate and the secret fraction: the former typically decreases when the final fidelity grows, whereas the latter increases when the final fidelity becomes larger. Our analysis allows to calculate secret key rates under the assumption of a single repeater chain with at most $2^k$ quantum memories per half station for respectively $k$ distillation rounds occurring strictly before
the swappings start. The use of additional memories when parallelizing or even multiplexing several such repeater chains as well as the use of additional quantum error detection or even correction will certainly improve these rates, but also render the experimental realization much more difficult.

The comparison of different protocols is highly subjective, as there are different experimental requirements and difficulties for each of them, therefore here we investigated the main aspects for every protocol separately.

The general type of quantum repeater is a kind of prototype for a quantum repeater based on the original proposal \cite{briegel_quantum_1998}. We have provided an estimate of the experimental parameters needed to extract a secret key and showed what the role of each parameter is. We have found that the requirement on the initial fidelity is not so strong if distillation is allowed. However, quantum gates need to be very good (errors of the order of $1\%$).

Further, we have studied the hybrid quantum repeater. This protocol permits to perform both the initial entanglement distribution and the entanglement swapping with high efficiencies. The reason is that bright light sources are used for communication and Cavity Quantum Electrodynamics (CQED) interactions are employed for the local quantum gates, making the swapping, in principle, deterministic. Using photon-number resolving detectors, we have derived explicit formulas for the initial fidelity and the probability of success for entanglement distribution. Furthermore, we have found the form of the states after entanglement swapping and entanglement distribution in the presence of gate errors. We have seen that finite detector efficiencies do not play a major role  regarding the generation probability. This permits to have high secret key rates in a set-up where it is possible to neglect imperfections of the detectors. By studying imperfect gates we found that excellent gates are necessary (errors of the order of $0.1\%$).

Finally, we have considered repeaters with atomic ensembles and linear optics. There exist many experimental proposals and therefore we have studied the scheme which is believed to be the fastest \cite{minar}. This scheme uses heralded qubit amplifiers for creating dual-rail encoded entanglement and entanglement swapping based on two-fold detection events. In contrast to the previous two schemes, the Bell measurement used for entanglement swapping is not able to distinguish all four Bell states. We have characterized all common imperfections and we have seen that using present technology, the performance of this type of quantum repeater in terms of secret key rates is only about one order of magnitude different from the corresponding ideal set-up. Thus, this scheme seems robust against most imperfections. These types of repeater schemes, as currently being restricted to linear optics, could still be potentially improved by allowing for additional nonlinear-optics elements. This
may render the entanglement swapping steps deterministic,
similar to the hybrid quantum repeater using CQED, and thus further enhance the secret key rates.

For  the protocols considered here, single-qubit rotations were assumed to be
perfect. Obviously, this assumption is not correct in any realistic situation. However, most of these single-qubit rotations can be replaced by simple bit flips of the classical outcomes which are used when the QKD protocol starts. Therefore, we see that in this case, specifically building a quantum repeater for QKD applications permits to relax the requirements on certain operations that otherwise must be satisfied for a more general quantum application, such as distributed quantum computation.

As an outlook our analysis can be extended in various directions: In our work we have considered standard quantum key distribution, in which Alice and Bob trust their measurement devices. To be more realistic, it is possible to relax this assumption and to consider device-independent quantum key distribution (DI-QKD) \cite{ekert1991quantum,PhysRevLett.98.230501,PhysRevLett.105.070501,PhysRevA.84.010304,PhysRevA.84.022325}. An analysis of the performance of long-distance DI-QKD can also be done using the methods that we developed in this paper.

A possible continuation of our work is the analysis of multiplexing \cite{Collins2007,sangouard_quantum_2011}. It has been shown that this technique has significant advantage in terms of the decoherence time required by the quantum memories. On the other hand it produces only a moderate increase of the repeater rate \cite{jiang_fast_2007, sangouard_quantum_2011, razavi_quantum_2009}. Possible future analyses include the effect on the secret key rate by distilling in all nesting levels \cite{Bratzik2012} or by optimizing the repeater protocol as done in Refs.~\cite{jiang_optimal_2007,van2009system}. Moreover, other repeater protocols which are based on quantum error correction codes \cite{jiang_quantum_2009, fowler.2010, nadja.2012} may help to increase the secret key rate.

\begin{acknowledgments}
The authors acknowledge financial support by the German Federal Ministry of Education and Research (BMBF, project QuOReP). The authors would like to thank the organizers and participants of the quantum repeater workshops (project QuOReP) held in Hannover and Bad Honnef in 2011 and 2012.  N.~K.B. and  P.~v.L. thank the Emmy Noether Program of the Deutsche Forschungsgemeinschaft for financial support.  S.~A. thanks  J.~Min\'a\v{r}  for enlightening discussions and insightful comments.
\end{acknowledgments}

 \appendix

\section{\label{sec:app:general}Additional material for the general framework}

\subsection{Generation rate with probabilistic entanglement swapping and distillation}

In this appendix, we give the derivation of \eref{eq:avgngen} in \sref{subsec:parameters} which describes the generation rate of entangled pairs per time unit $T_0$ with probabilistic entanglement swapping and distillation, i.e.,

\begin{equation}
\label{eq:avgngen2}
 \rrep^{\rm prob}= \frac{1}{T_0}\left(\frac{2}{3a}\right)^{N+k} P_{0}P_{ES}^{(1)}P_{ES}^{(2)}...P_{ES}^{(\nl)}\prod_{i=1}^{k}\pdis[i].
\end{equation}
In \cite{sangouard_quantum_2011} the formula has been derived only for the case without distillation and there it reads as follows,
\begin{equation}
\label{eq:avgngen3}
 \rrep^{\rm prob}= \frac{1}{T_0}\left(\frac{2}{3}\right)^{N} P_{0} P_{ES}^{(1)}P_{ES}^{(2)}...P_{ES}^{(\nl)},
\end{equation}
where $P_{0}$ is the probability to generate a pair for entanglement swapping. This formula was derived for small $P_0$.

In order to incorporate distillation into \eref{eq:avgngen3} we use the definition of the recursive probability $P_{L_0}[k]$ given in \eref{eq:PL0}, see \cite{NadjaHybrid}. It describes the generation probability of an entangled pair after $k$ rounds of purification. If we choose an appropriate $a<1$ such that  $Z_{1}(x)=\frac{3-2x}{x(2-x)}\geq\frac{3}{2x}a$ , we can rewrite $P_{L_0}[k]$:
\begin{eqnarray}
\label{eq:approxPL0}
 P_{L_0}[k]&=&\frac{\pdis[k]}{Z_{1}(P_{L_0}[k-1])}\leq\frac{2}{3a}\pdis[k]P_{L_0}[k-1]\nonumber \\
&=&\frac{2}{3a}\pdis[k]\frac{\pdis[k-1]}{Z_1(P_{L_0}[k-2])}\nonumber\\
&\leq&...\leq\left(\frac{2}{3a}\right)^{k}P_0\prod_{i=1}^{k}\pdis[i],
\end{eqnarray}
where in the last line $P_{L_0}[k]$ is a recursive formula. For deriving \eref{eq:avgngen2}, we replace in \eref{eq:avgngen3} $P_0$ by $P_{L_0}$ and we use \eref{eq:approxPL0}.

For the plots we have $L=600$ km and usually $\eta_d=0.9$ which leads to $P_{L_0}[k]\leq0.037$ and $a\leq0.994$.

\section{Additional material for the original quantum repeater\label{app:general-type}}

\subsection{Entanglement swapping\label{sec:brig:es}\label{app:ES}}

In this appendix we present the formulas of the state after entanglement swapping and the distillation protocol. Moreover, we bound also the role of dark counts in the entanglement swapping probability. 

\subsubsection*{The protocol}
We consider the total state $\rho_{ab}\otimes\rho_{cd}$. The entanglement swapping algorithm consists of the following steps:
\begin{enumerate}
 \item A CNOT is applied on system $b$ as source and $c$ as target.
 \item One output system is measured in the computational basis and the other one in the basis $\{\ket{+}:=\frac{\ket{H}+\ket{V}}{\sqrt{2}}, \ket{-}=\frac{\ket{H}-\ket{V}}{\sqrt{2}}\}$, obtained by applying a Hadamard gate.
 \item In the standard entanglement swapping algorithm, a single qubit rotation depending on the outcome of the measurement is performed. However, for the purpose of QKD it is not necessary to do this single-qubit rotation\footnote{Note that this step is different from \cite{briegel_quantum_1998}, where the single-qubit rotations were explicitly included.}. We propose that Bob collects the results of the Bell measurements, performs the standard QKD measurement and then he can apply a classical bit flip depending on the QKD measurement basis and on the Bell measurement outcomes.
\end{enumerate}
\subsubsection*{Formulas in the presence of imperfections}

We consider a set-up with two detectors $d_1$ and $d_2$. We associate the detection pattern of these two detectors with a two-dimensional Hilbert space, e.g $d_1={\rm click}, d_2={\rm no click}\Rightarrow\ket{H}=\ket{1_{d_1},0_{d_2}}$ and $d_1={\rm no click}, d_2={\rm click}\Rightarrow\ket{V}=\ket{0_{d_1},1_{d_2}}$ where $\{\ket{H}, \ket{V}\}$ are a basis of a two-dimensional Hilbert space which can be, for example, identified with horizontal and vertical polarizations of a qubit. We discard those events where there are no clicks or when both detectors click. If the detectors are imperfect, we may have an error in the detection of the quantum state. The POVM consists of two elements $\Pi_H\; (\Pi_V)$ which detect mode $\ket{H} (\ket{V})$:
\newcommand{\pd}{\pD}
\begin{eqnarray}
\label{eq:POVM}
 \Pi_H&:=\gamma\ket{H}\bra{H}+(1-\gamma)\ket{V}\bra{V},\\
 \Pi_V&:=\gamma\ket{V}\bra{V}+(1-\gamma)\ket{H}\bra{H},
\end{eqnarray}
with
\begin{eqnarray}
 \gamma&:=\frac{\etaD+\pd(1-\etaD)}{\etaD+2\pd(1-\etaD)},
\end{eqnarray}
where $\pd$ is the dark count probability of the detectors and $\etaD$ is their efficiency\footnote{The coefficient $\gamma$ can be calculated as follows: the POVM for having a click under the assumption of single-photon sources and imperfect detectors is given by
\begin{equation*}
E^{\rm (click)}=\pd \proj{0}+\left(1-(1-\pd)(1-\etaD)\right)\proj{1}
\end{equation*}
 and no click
\begin{equation*}
E^{\rm(no click)}=(1-\pd)\proj{0}+(1-\pd)(1-\etaD)\proj{1}.
\end{equation*}
When we say that the detector $a$ clicked, and $b$ did not click and we discard the vacuum events, and those where both detectors clicked, the POVM looks as follows:
\begin{eqnarray*}
&& E_a^{\rm(click)}\otimes E_b^{\rm (noclick)}\\
&&=\left(1-(1-\pd)(1-\etaD)\right)(1-\pd)\proj{1_a,0_b}\\
&&+\pd (1-\pd)(1-\etaD) \proj{0_a,1_b}.
\end{eqnarray*}
The trace is $(1-\pd)(\etaD+2\pd(1-\etaD))$, which is exactly the probability that we have this measurement. If we normalize this measurement and relate it to the POVM in \eref{eq:POVM}, we get $\gamma$.}.

The POVM above has been used also in \cite{briegel_quantum_1998,duer_quantum_1999}, however, the connection with the imperfections of the detectors was not made.

\newcommand{\bco}[2]{\lambda_{#1, #2}}

If we start with the states $\rho_{ab}=\rho_{cd}=A\proj{\phi^+}+B\proj{\phi^-}+C\proj{\psi^+}+D\proj{\psi^-}$, the resulting state after entanglement swapping between $a$ and $d$ is still a Bell diagonal state with coefficients of the form \cite{duer_quantum_1998}:
\newcommand{\bsum}[2]{\lambda_{[#1} \eta_{#2]}}
\label{eq:esfinstate}
\begin{widetext}
\begin{align}
 A'=&\frac{1-p_{G}}{4}+p_{G}\left[\gamma^2(A^2+B^2+C^2+D^2) + 2(1-\gamma)^2(AD+BC) +2\gamma(1-\gamma)(A+D)(C+B)\right],\nonumber\\
B'=&\frac{1-p_{G}}{4}+p_{G}\left[2\gamma^2(AB+CD)   + 2(1-\gamma)^2(AC+BD)+\gamma(1-\gamma)(A^2 +B^2 +C^2 +D^2 +2AD +2BC)\right],\nonumber\\
C'=&\frac{1-p_{G}}{4}+p_{G}\left[2\gamma^2(AC+BD)   + 2(1-\gamma)^2(AB+CD)+\gamma(1-\gamma)(A^2 +B^2 +C^2 +D^2 +2AD +2BC)\right],\nonumber\\
D'=&\frac{1-p_{G}}{4}+p_{G}\left[2\gamma^2(AD+BC)   + (1-\gamma)^2(A^2+B^2+C^2+D^2)+2\gamma(1-\gamma)(A+D)(B+C)\right],
\end{align}
\end{widetext}

and the probability to obtain the state above is equal to
\begin{equation}
\label{eq:briegpes}
 P_{ES}(\etaD, \pD):=\left((1-\pd)(\etaD+2\pd(1-\etaD))\right)^2,
\end{equation}
which can be interpreted as the probability that entanglement swapping is successful\footnote{This probability was derived by taking the probability of the measurement in the preceding footnote squared, as we need two coincident clicks for the Bell measurement.}. Note that $P(\eta, 0)=\eta^2$ and $P(1, 0)=1$ as we expect. When we consider dark counts $p_{\rm dark}<10^{-5}$, then these are negligible as $(P_{ES}(0.1,10^{-5})/(P_{ES}(0.1,0)))^N<1.03^N$, so the impact on the secret key rate is minimal.
Note that we open the gates only for a short time window, which is the interval of time where we expect the arrival of a photon. The dark count probability $\pdark$ represents the probability that in the involved time window the detector gets a dark count.

\subsection{Distillation}

\subsubsection*{The protocol}
We assume that Alice and Bob hold two Bell diagonal states $\rho_{a_1,b_1}$ and $\rho_{a_2,b_2}$. The algorithm is the following:
\begin{enumerate}
 \item In the computational basis, Alice rotates her particles by $\frac{\pi}{2}$ about the $X$-axis, whereas Bob applies the inverse rotation ($-\frac{\pi}{2}$) on his particles.
\item Then they apply on both sides a CNOT operation, where the states $a_1\; (b_1)$ serve as source and $a_2\; (b_2)$ as target.
\item The states corresponding to the target are measured in the computational basis. If the measurement results coincide, the resulting state $\rho_{a_1, b_1}$ is a purified state; otherwise, the resulting state is discarded. Therefore, this entanglement distillation scheme is probabilistic.
\end{enumerate}

\subsubsection*{\label{sec:appDistill}Formulas in the presence of imperfections}
Given a Bell diagonal state with the following coefficients
\begin{equation}
\rho_{ab}=A\proj{\phi^+}+B\proj{\phi^-}+C\proj{\psi^+}+D\proj{\psi^-},
\label{eq:bell}
\end{equation}

the coefficients transform according to the following map \cite{deutsch1996quantum}:
\begin{eqnarray}
A'&=& \frac{1}{\pdis}\left(A^2+D^2\right),\\
B'&=&\frac{1}{\pdis}\left(2AD\right),\\
C'&=& \frac{1}{\pdis}\left(B^2+C^2\right),\\
D'&=&\frac{1}{\pdis}\left(2BC\right),
\end{eqnarray}
where $\pdis$ is the probability that the measurement outcomes are both the same for Alice and Bob, and thus the probability of successful distillation is:
\begin{equation}
\pdis[k]=\left(A_{k-1} +D_{k-1}\right)^2+\left(B_{k-1}+C_{k-1}\right)^2.
\label{eq:pevenD}
\end{equation}

Including the gate quality $p_G$, these formulas change to \cite{duer_quantum_1998}:

\begin{equation}
\pdis[k]=\frac{1}{2}\left\{1+p_G^2\left(-1+2A_{k-1} +2D_{k-1}\right)^2\right\}.
\label{eq:pevenDe}
\end{equation}

with
\begin{widetext}
\begin{eqnarray*}
A'&=&\left[1+p_G^2\left( (A-B-C+D) (3 A+B+C+3 D)+4 (A-D)^2\right)\right]/(8 \pdis),\\
B' &=&\left[1-p_G^2 \left(A^2+2 A (B+C-7 D)+(B+C+D)^2\right)\right]/(8 \pdis),\\
C' &=&\left[1+p_G^2 \left(4 (B-C)^2-(A-B-C+D) (A+3 (B+C)+D)\right)\right]/(8 \pdis),\\
D' &=& \left[1-p_G^2 \left(A^2+2 A (B+C+D)+B^2+2 B (D-7 C)+(C+D)^2\right)\right]/(8 \pdis).
\end{eqnarray*}
\end{widetext}

\section{\label{sec:app:hybrid}Additional material for the hybrid quantum repeater}
In this appendix we derive the formula for  successful entanglement generation when PNRD are used for the measurements. Moreover, we present the formulas for the states after entanglement swapping and entanglement distillation.
\subsection{\label{sec:app:gen}Entanglement generation}
The total state before the detector measurements is described by \cite{azuma}
\begin{widetext}
\begin{eqnarray}
\rho_{AB,b_3,b_5}=&p\left\{\left[\ket{0}_{b_3}(\ket{00}_{AB}\ket{\beta}_{b_5}+\ket{11}_{AB}\ket{-\beta}_{b_5})/2+\ket{0}_{b_5}(\ket{01}_{AB}\ket{-\beta}_{b_3}+\ket{10}_{AB}\ket{\beta}_{b_3})/2\right]\times H.c.\right\}+\nonumber\\
&(1-p)\left\{\left[\ket{0}_{b_3}(\ket{00}_{AB}\ket{\beta}_{b_5}-\ket{11}_{AB}\ket{-\beta}_{b_5})/2+\ket{0}_{b_5}(\ket{01}_{AB}\ket{-\beta}_{b_3}-\ket{10}_{AB}\ket{\beta}_{b_3})/2\right]\times H.c.\right\},
\label{app:instate}
\end{eqnarray}
\end{widetext}

where $H.c.$ stays for the Hermitian conjugate of the previous term, $A$ ($B$) represents the qubit at Alice's (Bob's) side, $b_3$ is the coherent-state mode arriving at the detector $D_1$, $b_5$ is the coherent-state mode arriving at the detector $D_2$, and $\beta=i\sqrt{2\eta_t}\sin{(\theta/2)}$ (see figure~\eref{fig:hqr}). The probability of error caused by photon losses in the transmission channel is given by $(1-p)$, with $p=(1+e^{-2(1-\eta_t)\alpha^2\sin^2{(\theta/2)}})/2$. It is possible to observe from \eref{app:instate} that whenever Bob detects a click in either one of the detectors $D_1$ or $D_2$, an entangled state has been distributed between qubits $A$ and $B$.

We discuss in the following the case that $D_1$ and $D_2$ are imperfect PNRD (see \eref{eq:POVMPNRD}). When detector $D_1$ does not click and $D_2$ clicks, the resulting state $\rho_{AB}$ is then given by
\begin{equation}
\label{eq:rhoabhybridapp}
\rho_{AB}=\frac{\mbox{tr}_{b_3b_5}(\Pi_{b_3}^{(0)}\Pi_{b_5}^{(n)}\rho_{AB,b_3,b_5})}{\mbox{tr}(\Pi_{b_3}^{(0)}\Pi_{b_5}^{(n)}\rho_{AB,b_3,b_5})},
\end{equation}
with $n>0$. The same result up to local operations can be obtained in the opposite case (a click in detector $D_1$ and no click in detector $D_2$).

Depending on the outcome of the detector, a local operation maybe applied to change the resulting state into the desired state. In this way, if the outcome is an even number, nothing should be done, otherwise a $Z$ operation should be applied. Following this, the resulting state can be written as
\begin{equation}
\rho=F_0\ket{\phi^+}\bra{\phi^+}+(1-F_0)\ket{\phi^-}\bra{\phi^-},\nonumber
\end{equation}
where
\begin{eqnarray}
F_0=&\frac{(\bra{00}_{AB}+(-1)^n\bra{11}_{AB})}{\sqrt{2}}\rho_{A,B}\frac{(\ket{00}_{AB}+(-1)^n\ket{11}_{AB})}{\sqrt{2}}\nonumber\\
=&\frac{1+e^{-2(1+\eta_t(1-2\eta_d))\alpha^2\sin^2(\theta/2)}}{2}.
\end{eqnarray}
The probability of success is calculated by adding all successful events, and is given by
\begin{equation}
P_0=\sum_{n=1}^{\infty}\mbox{tr}(\Pi_{b_3}^{(0)}\Pi_{b_5}^{(n)}\rho_{AB,b_3,b_5}+\Pi_{b_5}^{(0)}\Pi_{b_3}^{(n)}\rho_{AB,b_3,b_5}).
\end{equation}

Combining \eref{app:instate} and \eref{eq:POVMPNRD} we obtain \eref{initial.p0}.

\subsection{\label{sec:app:swap}Entanglement swapping}

The initial states used in the swapping operation are a full rank mixture of the Bell states, $\rho_0:=A\ket{\phi^+}\bra{\phi^+}+B\ket{\phi^-}\bra{\phi^-}+C\ket{\psi^+}\bra{\psi^+}+D\ket{\psi^-}\bra{\psi^-}$. After the connection, the resulting state will remain in the same form, $A'\ket{\phi^+}\bra{\phi^+}+B'\ket{\phi^-}\bra{\phi^-}+C'\ket{\psi^+}\bra{\psi^+}+D'\ket{\psi^-}\bra{\psi^-}$, but with new coefficients:
\begin{widetext}
\begin{align}
A'&=2 B C + 2 A D +
 2 [-2 B C + A (B + C - 2 D) + (B + C) D] p_G + (A - B - C + D)^2 p_G^2,\nonumber\\
B'&=2 A C + 2 B D + [A^2 + (B + C)^2 - 4 B D + D^2 +
     2 A (-2 C + D)] p_G - (A - B - C + D)^2 p_G^2,\nonumber\\
C'&=2 A B + 2 C D + [A^2 + (B + C)^2 - 4 C D + D^2 +
    2 A (-2 B + D)] p_G - (A - B - C + D)^2 p_G^2,\nonumber\\
D'&=A^2 + B^2 + C^2 + D^2 -
 2 [A^2 + B^2 + C^2 - A (B + C) -(B + C) D + D^2] p_G  + (A - B - C +
    D)^2 p_G^2.
\label{coeff}
\end{align}
\end{widetext}

It is possible to see that $A'+B'+C'+D'=1$, such that even for the case of imperfect connection operations, the swapping occurs deterministically.

\subsection{\label{sec:app:dist}Entanglement distillation}
We calculated also the effect of the gate error in the distillation step. Starting with two copies of states in the form of $\rho_0:=A\ket{\phi^+}\bra{\phi^+}+B\ket{\phi^-}\bra{\phi^-}+C\ket{\psi^+}\bra{\psi^+}+D\ket{\psi^-}\bra{\psi^-}$, the resulting state after one round of distillation is given by $A'\ket{\phi^+}\bra{\phi^+}+B'\ket{\phi^-}\bra{\phi^-}+C'\ket{\psi^+}\bra{\psi^+}+D'\ket{\psi^-}\bra{\psi^-}$, where
\begin{widetext}
\begin{align}
A'&=\frac{1}{\pdis}\left(D^2 + A^2 [1 + 2 (-1 + p_G) p_G]^2 -
 2 A (-1 + p_G) p_G [C + 2 D+ 2 (B - C - 2 D) p_G   +2 (-B + C + 2 D) p_G^2]\right.\nonumber\\
&\left.-2 D (-1 + p_G) p_G \{-2 D - 2 (C + D) (-1 + p_G) p_G+ B [1 + 2 (-1 + p_G) p_G]\}\right),\nonumber\\
B'&=\frac{1}{\pdis}\left[-2 \bm{(}D (-1 + p_G) p_G (C + D + 2 B p_G - 2 C p_G - 2 D p_G - 2 B p_G^2 + 2 C p_G^2 +
      2 D p_G^2)+A^2 p_G (-1 + 3 p_G - 4 p_G^2 + 2 p_G^3)\right.\nonumber\\
&\left.-A \{D (1 - 2 p_G + 2 p_G^2)^2 - (-1 + p_G) p_G [-2 C (-1 + p_G) p_G +
         B (1 - 2 p_G + 2 p_G^2)]\}\bm{)}\right],\nonumber\\
C'&=\frac{1}{\pdis}\left(B^2 (1 - 2 p_G + 2 p_G^2)^2 -
 2 B (-1 + p_G) p_G [-2 A (-1 + p_G) p_G + D (1 - 2 p_G + 2 p_G^2) +
    C (2 - 4 p_G + 4 p_G^2)] \right.\nonumber\\
&\left.+C \{C (1 - 2 p_G + 2 p_G^2)^2-
    2 (-1 + p_G) p_G [-2 D (-1 + p_G) p_G+ A (1 - 2 p_G + 2 p_G^2)]\}\right),\nonumber\\
D'&=\frac{1}{\pdis}\left\{-2 (C (-1 + p_G) p_G (C + D + 2 A p_G - 2 C p_G - 2 D p_G - 2 A p_G^2 + 2 C p_G^2 +
      2 D p_G^2)  +B^2 p_G (-1 + 3 p_G- 4 p_G^2 + 2 p_G^3)\right. \nonumber\\
&\left.-B \{C (1 - 2 p_G + 2 p_G^2)^2- (-1 + p_G) p_G [-2 D (-1 + p_G) p_G +
         A (1 - 2 p_G + 2 p_G^2)]\})\right\},
\label{coeffpur}
\end{align}
\end{widetext}
$\pdis$ is the distillation probability of success and is given by
\begin{eqnarray}
\pdis=&&(B + C)^2 + (A + D)^2 - 2 (A - B - C + D)^2 p_G\nonumber\\
&& +2 (A - B - C + D)^2 p_G^2.
\label{ppur}
\end{eqnarray}
For the case of $p_G=1$, \eref{coeffpur} and \eref{ppur} are in accordance with \cite{deutsch1996quantum}.
\bibliography{allbib}
\end{document}